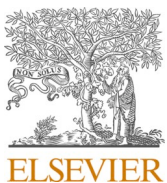
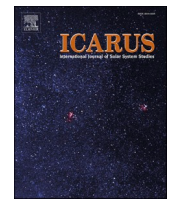

# The impact of lake shape and size on lake breezes and air-lake exchanges on Titan

Audrey Chatain [a,b,*], Scot C.R. Rafkin [b], Alejandro Soto [b], Enora Moisan [b,c], Juan M. Lora [d], Alice Le Gall [e,f], Ricardo Hueso [a], Aymeric Spiga [g]

[a] *Departamento de Física Aplicada, Escuela de Ingeniería de Bilbao, Universidad del País Vasco/Euskal Herriko Unibertsitatea (UPV/EHU), Bilbao, Spain*
[b] *Department of Space Studies, Southwest Research Institute (SwRI), Boulder, CO, USA*
[c] *ENS de Lyon, University of Lyon, CNRS, LGL-TPE, 46 Allée d'Italie, F-69007 Lyon, France*
[d] *Department of Earth and Planetary Sciences, Yale University, New Haven, CT, USA*
[e] *Laboratoire Atmosphères Observations Spatiales/Institut Pierre-Simon Laplace (LATMOS/IPSL), Université Paris-Saclay, Université de Versailles Saint-Quentin-en-Yvelines (UVSQ), Sorbonne Université, Centre National de la Recherche Scientifique (CNRS), Guyancourt, France*
[f] *Institut Universitaire de France (IUF), Paris, France*
[g] *Laboratoire de Météorologie Dynamique/Institut Pierre-Simon Laplace (LMD/IPSL), Centre National de la Recherche Scientifique (CNRS), Sorbonne Université, Paris, France*

ABSTRACT

Titan, the largest moon of Saturn, has many lakes on its surface, formed mainly of liquid methane. Like water lakes on Earth, these methane lakes on Titan likely profoundly affect the local climate. Previous studies (Rafkin and Soto, 2020; Chatain et al., 2022) showed that Titan's lakes create lake breeze circulations with characteristic dimensions similar to the ones observed on Earth. However, such studies used a model in two dimensions; this work investigates the consequences of the addition of a third dimension to the model. Our results show that 2D simulations tend to overestimate the extension of the lake breeze over the land, and underestimate the strength of the subsidence over the lake, due to divergence/convergence geometrical effects in the mass conservation equations. In addition, 3D simulations including a large scale background wind show the formation of a pocket of accelerated wind behind the lake, which did not form in 2D simulations. An investigation of the effect of shoreline concavity on the resulting air circulation shows the formation of wind currents over peninsulas. Simulations with several lakes can either result in the formation of several individual lake breeze cells (during the day), or the emergence of a large merged cell with internal wind currents between lakes (during the night). Simulations of several real-shaped lakes located at a latitude of 74°N on Titan at the autumn equinox show that larger lakes trigger stronger winds, and that some sections of lakes might accumulate enough methane vapor to form a thin fog. Additionally, we adapted the Turbulent Kinetic Energy closure scheme of the model to better represent the extremely low turbulence at the surface of Titan, of $2\,10^{-4}$ $m^2.s^{-2}$ above the land, and inferior to $3\,10^{-5}$ $m^2.s^{-2}$ above the lake. The addition of a third dimension, along with adjustments in the parametrizations of turbulence and subsurface land temperature, results in a reduction in the magnitude of the average lake evaporate rate, namely to ~6 cm/Earth year.

## 1. Introduction

Titan, Saturn's largest moon, is the only known place except the Earth to host large reservoirs of liquids on its surface. These reservoirs are mostly made of methane, with ethane and possibly other hydrocarbons mixed in (e.g., Mastrogiuseppe et al., 2014, 2016, 2019; Mitchell et al., 2015; Stofan et al., 2007). Interestingly, these reservoirs have a large diversity of sizes and shapes. The largest ones, ranging from 380 to 1170 km in length, form seas. The Cassini mission also observed several hundred smaller lakes, with lengths ranging from a few kilometers to 240 km (Hayes, 2016; Hayes et al., 2008). Some lakes have smooth round shapes, while others have very complex shorelines. Some are isolated, while others are grouped in clusters forming a *lake district* (MacKenzie et al., 2019a).

* Corresponding author at: Departamento de Física Aplicada, Escuela de Ingeniería de Bilbao, Universidad del País Vasco/Euskal Herriko Unibertsitatea (UPV/EHU), Bilbao, Spain.
*E-mail address:* audrey.chatain@latmos.ipsl.fr (A. Chatain).

Studies of lakes on Earth show that large lakes profoundly affect the local climate because of their different thermal inertia compared to land, surface cooling associated with evaporation, their lower surface roughness and albedo compared with land, and other factors (e.g., Changnon Jr and Jones, 1972; Crosman and Horel, 2010). In particular, lakes can induce local winds due to temperature differences with the nearby land. These thermal contrasts are created because land terrains have typically lower thermal inertia than lakes, and because lakes can cool due to evaporation. Lake and land breezes can thus be formed with insolation variations. With the objective to understand and forecast sea and lake breezes, many modelling works have investigated the effect of the lake shapes, sizes and background conditions on the induced winds (for a review, see Crosman and Horel, 2010). In particular, the shoreline convexity has been shown to affect the circulation by distorting the lake breeze divergence field (e.g., McPherson, 1970), and larger lakes have been found to force a stronger circulation (e.g., Boybeyi and Raman, 1992). In addition, the necessity to include the large-scale background wind in 3D mesoscale simulations to reproduce sea breeze observations has been shown in various studies (e.g., Drobinski et al., 2006), as has the importance of including neighboring lakes whose breeze circulations could locally interfere in complex ways (e.g., Daggupaty, 2001).

As lakes strongly affect the weather and hydrological cycle on Earth, similar questions arise on Titan. Nevertheless, the lake-induced circulations could be quite different from what happens on the Earth due to the very different conditions at Titan's surface. While the pressure of 1.5 bar at the surface is similar to the Earth, the temperature is much colder (~92 K) with very limited diurnal and seasonal variations (e.g., Cottini et al., 2012; Jennings et al., 2019). The compositions of the atmosphere, the lakes and the land are also different, with methane vapor and liquid instead of water, and a deposit of photochemically-produced organics over an icy bedrock instead of silicate rocks, overlain by soil and vegetation. As these materials have very different properties, and radiative heating and temperatures are much lower on Titan, we would expect lake-induced circulations to be quantitatively different on Titan, forming a slow-motion version of the dynamics observed in Earth's lake breezes.

On Titan, the origin of the atmospheric methane remains a puzzle. Its high concentration at the surface (~5%, Niemann et al., 2010) could have various sources, such as the evaporation of lakes and seas, the evaporation of highly humid lands, or possibly large surface or subsurface catastrophic events (like cryovolcanism or resurgence of a subsurface alkanofer) (Faulk et al., 2020; Griffith et al., 2008; Mitchell and Lora, 2016; Tobie et al., 2006; Turtle et al., 2018). Previous works have intended to constrain the evaporation from the lakes. A first analytic 1D model (Mitri et al., 2007) estimated the lake evaporation rate to be 3–10 m/Earth year, though large-scale models indicated much lower values (e.g. Lora et al., 2015; Mitchell, 2008, 2012; Newman et al., 2016; Schneider et al., 2012). More recent studies (Chatain et al., 2022; Rafkin and Soto, 2020) showed, with a 2D mesoscale model including a lake, that a lake-breeze circulation is created and maintained above the lake in all insolation conditions, which is different from lakes on the Earth where lake-breezes appear mostly in summer during the day. The evaporation of methane from the lake then forms a cold and moist stable layer above the lake, which inhibits methane turbulent mixing and substantially slows down the lake evaporation to 0.2–0.8 m/Earth year.

To improve our understanding of lake-induced meteorological effects on Titan, this work follows the lessons learned from studies on Earth's lakes and investigates the impact of lake shape and size on the lake-breeze structure and air-lake exchanges. We study the effect of the addition of the third spatial dimension in comparison to prior studies (Chatain et al., 2022; Rafkin and Soto, 2020) by first using an idealized circular lake (Section 3). Then, additional simulations using realistic Titan shorelines explore the effect of shoreline concavity, proximity to other lakes (Section 4), and lake size (Section 5) on evaporation and the local atmospheric circulation.

## 2. Model description

### 2.1. Model description and improvements

We use the Titan mesoscale model *mtWRF* (for "mesoscale Titan WRF"), which simulates the atmospheric circulation over a methane lake surrounded by land, initially described in Rafkin and Soto (2020). We also use the developments described in Chatain et al. (2022) that included a grey radiative scheme describing a broadband solar radiation varying diurnally, and a broadband infrared channel. This work also uses double precision computation (64 bit), which negates the need for tendency accumulators utilized in all the prior mtWRF studies to adequately capture the very small temperature variations obtained in the model. Furthermore, the minimum allowable turbulent kinetic energy within the subgrid turbulence model is decreased (to $3 \cdot 10^{-5}$ m$^2$.s$^{-2}$) to better represent Titan's surface conditions, which are extremely stable compared to the Earth ($> 0.1$ m$^2$.s$^{-2}$). The sensitivity on the chosen value is detailed in Appendix 1. It does not affect the circulation, but it has a non-negligible effect on the methane mixing in the first meters above the lake. This can affect the relative humidity just above the lake (by a few percent), the lake evaporation (by 20–30%), and the lake temperature (by ~0.2 K). Additionally, a new method is used to prescribe the lower boundary condition of the subsurface land temperature (see details in Appendix 2).

Surface physical properties over land are homogeneous in the model and there is no topography. Sensible heat exchange is permitted with the air, but there is no latent heat or vapor exchange with the land surface. We note that moist terrains are likely to be present on Titan around lakes and potentially evaporate some methane, though with a lower efficiency than lakes. A future work is planned on the parametrization of such methane exchanges between the land and the air in further developments of the model. The model does not include a microphysics scheme. However, atmospheric saturation conditions are never reached in the simulations presented here, and condensation and precipitation are therefore not likely to happen.

The surface roughness length is changed to 0.4 cm, compared to the default 40 cm used in the previous works, because a smaller roughness is more appropriate for Titan's plains (Lopes et al., 2016; Nelli et al., 2020; Tokano, 2005). The surface roughness length is an empirical parameter corresponding to the altitude above the surface at which the horizontal wind is null in the logarithmic velocity profile equation of the horizontal wind (Arya, 2001). The surface roughness length is proportional to the mean height of the roughness elements, and their density (Garratt, 1977; Lettau, 1969). The surface infrared emissivity is updated to 0.95 over land and 0.99 over the lake (which were both equal to 0.9 in previous versions), based on Solomonidou et al. (2020) and Le Gall et al. (2016). The latter work provides a microwave emissivity while the model requires an infrared emissivity but the difference should be very small over the lakes. As in Chatain et al. (2022), the thermal inertia is fixed to 601 Jm$^{-2}$K$^{-1}$s$^{-0.5}$ on the land (assumed as a "plain" unit in MacKenzie et al., 2019b), and the surface Bond albedo is 0.3 over the land, and 0.1 over the lake.

Air temperature, humidity, and the large scale horizontal wind are initialized from a single profile adapted from a recent simulation (Lora et al., 2022) with the general circulation model (GCM) TAM (Titan Atmospheric Model, Lora et al., 2015) at a latitude of 74°N, at the autumn equinox, over a lake. The system then freely evolves from the initial conditions with periodic boundary conditions. Other TAM soundings over different lakes at 74°N are nearly identical. The vertical resolution of the GCM is insufficient to resolve the structures over the lowest few hundred meters. To better constrain the wind structure close to the surface, a 2D mtWRF simulation was run over land using the low-resolution GCM wind as an input. The land friction slows down the wind close to the surface, and the stationary wind solution obtained after a few tsols (Titan days) is used as the initial wind condition. The final inputs used for our simulations are shown in Fig. 1. The profiles simulated by TAM at 74°N at the autumn equinox are notably different from the Huygens profile used in prior mtWRF studies and are presumably more representative of the conditions near high latitude lakes (the Huygens landing site is at a latitude of 10.6°S).

*2.2. Simulation configurations*

Previous simulations (Chatain et al., 2022; Rafkin and Soto, 2020) used 2D configurations and provided insights on the meteorological processes operating in the vicinity of methane lakes. To obtain additional quantitative results, we run the model with three spatial dimensions and use lake shorelines observed by Cassini radar images of Titan (Hayes, 2016). In this study, we select lakes located at a similar latitude (~74°N) to isolate the effects of shoreline concavity, proximity to other lakes, and lake size on the lake breeze circulation, and to remove any effect related to latitudinal variations of insolation or climate. The modelling domain is 201 × 201 points in the horizontal with 50 vertical levels extending from the surface to 15 km. The top 5 km contain a gravity wave damping layer. The vertical elevation of the model domain is more than sufficient since the lake breeze within the planetary boundary layer is at most 1 km (see e.g. the first figure of Section 3.1). Simulations with lakes of ~50 km horizontal dimension use a horizontal grid spacing of 2 km. Simulations with larger lakes use a grid spacing of 4 km. A comparison of model simulations using 2 km and 4 km grid spacing with an otherwise identical configuration confirms that both cases lead to similar results (see Appendix 3). The lake is modelled by a single slab layer of pure liquid methane, which physically represents the lake mixed layer. The layer is vertically homogeneous and its temperature instantaneously responds to a net change in energy flux. The slab model does not account for horizontal heat conduction or convection, and neither for exchanges with the deeper layers of the lake. The lake mixed layer is fixed to 1 m so that the lake breeze solution quickly stabilizes. Chatain et al. (2022) demonstrated that deeper mixed layers give similar results after a longer stabilization time. The simulation time step is 63.6 s. The diurnal variation of variables is repeatable after 2 tsols (Titan days) and the runs are terminated after 4 tsols.

Table 1 summarizes key domain parameters for the 10 configurations considered in this study. Each configuration is tested with and without an initial background wind to isolate that effect. A 2D configuration (simulations b and B) is also tested to directly investigate the impact of adding the third dimension.

**Table 1**
Parameter settings for all the simulations.

| Lake configuration | Horizontal resolution (km) | Simulation name (no wind) | Simulation name (with wind) |
|---|---|---|---|
| Circular, 50 km diameter | 2 | a | A |
| 50 km large, in 2D | 2 | b | B |
| Feia Lacus | 2 | c | C |
| Abaya Lacus | 2 | d | D |
| Oneida Lacus | 2 | e | E |
| Oneida Lacus region | 2 | f | F |
| Circular, 50 km diameter | 4 | g | G |
| Circular, 100 km diameter | 4 | h | H |
| Bolsena Lacus | 4 | i | I |
| Jingpo Lacus | 4 | j | J |

**3. The reference simulation: a circular lake**

*3.1. A realistic divergence field with 3D simulations*

Simulation "a" assuming a circular 50 km lake serves as the reference simulation. This section discusses the simplified case without background wind. Fig. 2 compares the wind circulations in 2D (simulation b) to this reference. In both cases, a lake breeze is present, blowing at the surface from the cold lake to the warmer land. The boundary separating the two different types of air (from the land and from the lake breeze) is called the "front" of the lake breeze. At all times, there is a strong vertical wind at the lake breeze front. Then the lake breeze cell is closed by a reverse horizontal wind at altitude and subsidence over the lake. This breeze has diurnal variations, previously described for Titan in Chatain et al. (2022), and similarly observed on Earth (for a review see Crosman and Horel, 2010). On Titan, during the night and at low insolation, the lake breeze extends far over land, while higher insolation conditions during the day create turbulent convection that mixes and destroys the thermal gradient along the lake breeze front and reduces the lake breeze penetration distance over the land (i.e. the distance of the front to the lake). The circulation has a similar vertical extent in the 3D and 2D cases. However, the horizontal extent is notably different. In the 2D case, the lake breeze circulation goes further inland (to a maximum of 150 km from the lake center, compared to 100 km in 3D). Except close to the lake, the 2D case overestimates the horizontal and vertical winds over the land. As an example, the daytime surface wind over the land at 50 km from the center of the lake reaches values more than twice overestimated in 2D compared to 3D (0.08 m/s compared to 0.03 m/s). Relatedly, the subsidence over the lake in 2D is underestimated by a factor of 2.

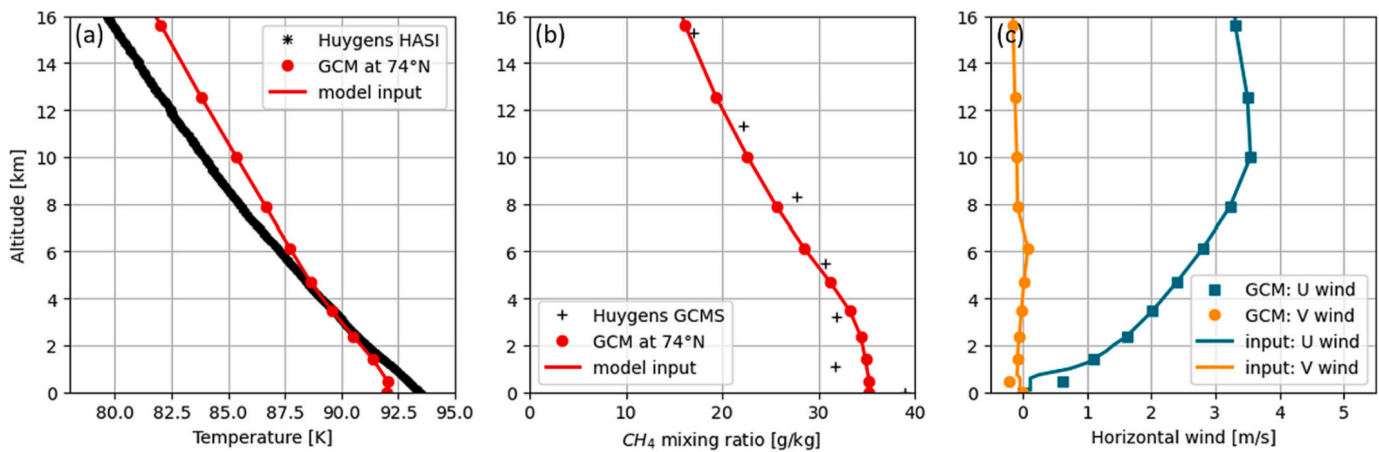

**Fig. 1.** Temperature (a), methane mixing ratio (b), and horizontal wind (c) from Huygens HASI (Fulchignoni et al., 2005) and GCMS (Niemann et al., 2010) measurements at 10.6°S just after the southern summer solstice, from the TAM GCM at 74°N at the autumn equinox (Lora et al., 2022), and used as inputs for our model.

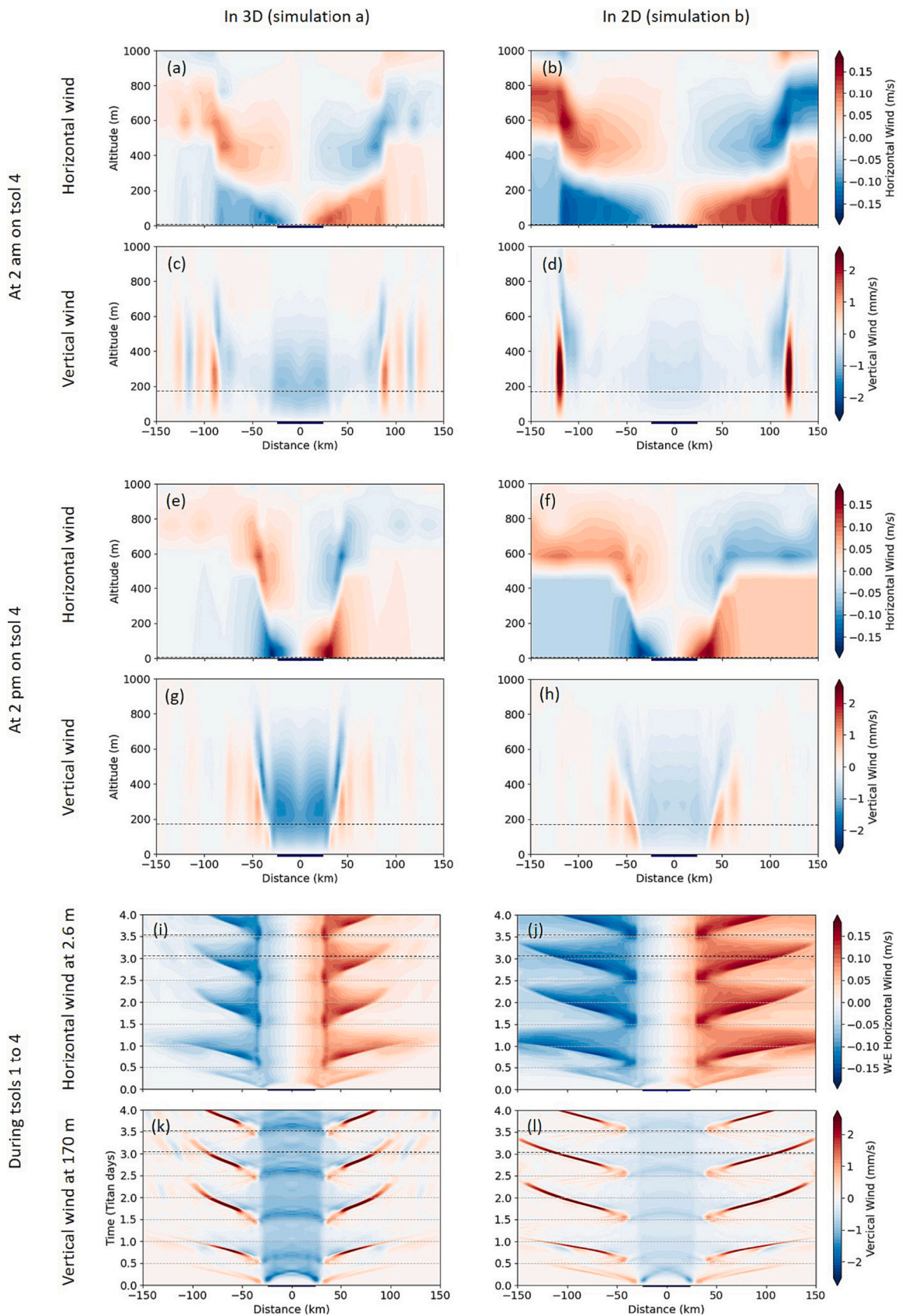

**Fig. 2.** Horizontal and vertical winds as function of altitude, distance to the lake center and time. Comparison of circulations obtained in 3D (simulation a) and 2D (simulation b) for a 50 km diameter lake. Simulations start at midnight. The dark blue bars below the plots indicate the position and size of the lake. The black dashed lines indicate the position of the slices plotted with the different axes (with altitude from a to h, with time from i to l). (For interpretation of the references to colour in this figure legend, the reader is referred to the web version of this article.)

These differences are explained by the fact that in the 3D simulation the lakeshore has a curvature that creates horizontal divergence of the lake breeze. Running a 2D simulation is equivalent to running a 3D simulation with a rectangular lake that is infinitely long along the y dimension (as shown in Fig. 3), with a lakeshore that has no curvature at all. The effect of lakeshore curvature on the induced lake breeze has been previously observed and modelled for Earth lakes (Crosman and Horel, 2010). Boybeyi and Raman (1992) show that the effect of lakeshore curvature on the lake breeze structure and intensity is nonlinear. Lakes with a more pronounced curvature are associated with a more divergent breeze at the surface. That is to say, in an ideal case without air density variations and without any air flux coming from above, the horizontal wind intensity at the surface decreases with distance to the lake by simple divergence and mass continuity. This ideal case would give the equation: $U(r).r = U_0.R$, with $U_0$ the horizontal wind at the shore of a round lake of radius $R$, and $U(r)$ the horizontal wind at the distance $r$ from the lake center. In the case with no curvature (like in a 2D simulation), both $R$ and $r \to \infty$, and we would simply have $U(r) = U_0$, $\forall r$. These simplistic equations only provide rough guidance, and cannot be used as such in real cases because of the higher complexity. First, the air density varies with distance from the lake as the temperature increases. Second, vertical winds (downdraft and updraft) affect mass continuity at the surface. Third, lakes with a more pronounced curvature (e.g., in the 3D case here) are smaller than lakes with a less pronounced curvature (or no curvature at all like in the 2D case), and lake size also affects the strength of the lake breeze, with stronger winds created by larger lakes (this point is detailed further in Section 5). In conclusion, the change from 2D to 3D on the horizontal winds cannot be described by simple scaling equations and instead requires 3D simulations. In addition, the response of the vertical wind demanded by mass continuity changes between 2D and 3D. Mass continuity and geometrical divergence make the vertical wind in the updraft over the land weaker in 3D than in 2D, especially when the updraft is far from the lakeshore (i.e., at night). The reverse effect occurs with the subsidence over the lake: due to flux convergence and mass continuity, the downward vertical wind is stronger in 3D than in 2D.

The difference in the winds between 2D and 3D results in differences in air-lake exchange (Fig. 4). In particular, variations in the circulation have repercussions on the evaporation of the lake: a weaker subsidence over the lake in 2D leads to reduced dry air subsidence over the lake. As a result, the humidity is higher, and the evaporation lower compared to 3D. The evaporation obtained in 2D is underestimated by ∼15% compared to the 3D results (compare the light blue curves on Fig. 4c,i), which is non-negligible when computing the estimated release of methane from the lakes to the atmosphere. As a direct consequence of the decreased evaporation, the cooling of the lake through evaporation is less efficient in 2D and the lake temperature is overestimated by 0.1 K (Fig. 4f,l). This could be a significant variation on Titan (where surface temperature diurnal variations are at most 1–1.5 K, Cottini et al., 2012) and could potentially affect lake chemistry (Kumar and Chevrier, 2020; Steckloff et al., 2020) and currents. Although the evaporation of methane is on average more efficient over the lake in the 3D simulation, the mixing of the moist air (in the sense of methane humidity) coming from the lake with the drier air over the land, which is done by wind-driven advection and small scale turbulence, is also more efficient in 3D. This is a consequence of the 3D geometric divergence of the wind flow around the lake. In more details, the vapor methane mixing ratio is very similar at the shore in the 2D and 3D cases, but in the 2D case, it attenuates less strongly with distance from the lake (compare the orange dash-dotted d = 34 km lines on Fig. 4e,k). As the cold moist air propagates further over land in 2D, it cools the land through sensible heat flux (not shown). Therefore, the 2D case underestimates the temperature of the land at some distance from the lake (Fig. 4f,l).

In conclusion, 2D simulations result in a qualitatively similar lake breeze circulation as 3D simulations, but with an overestimated horizontal extent, an overestimated horizontal wind over the land, and an underestimated subsidence over the lake. As a direct consequence from these modifications in the circulation, the evaporation of the lake is underestimated and the temperature of the lake is overestimated. Therefore, we demonstrate here that it is necessary to run simulations in 3D to obtain better estimates of the wind, and at a lower extend of the evaporation and lake temperature.

### 3.2. Typical effect of a background wind

On Earth, background winds of 2–4 m/s are strong enough to significantly affect the lake breeze circulation, and background winds above 6–11 m/s even completely remove the lake breeze (Crosman and Horel, 2010). On Titan, background winds are expected to be weaker compared to Earth. Titan GCMs suggest winds in the order of 3.5 m/s at 10 km altitude (Lora et al., 2019), 1 m/s at 1200 m, and 0.12 m/s at a few hundred meters (Fig. 1c). Lake breeze winds are expected to be much smaller on Titan than on Earth, with a maximum of ∼0.2 m/s (this study), compared to ∼5 m/s on Earth. In this section, we look into the effect of the background wind on the 3D Titan lake breeze circulation and investigate if lake breezes stand through the relatively high background wind expected on Titan.

The reference simulation done without background wind (simulation a) is compared to the same simulation done with the background wind derived from TAM (simulation A) (see Fig. 5 and Fig. 6). As observed on Earth with moderate background winds (Crosman and Horel, 2010) and on Titan with previous 2D simulations (Chatain et al., 2022; Rafkin and Soto, 2020), the lake breeze is still present, albeit deformed by the presence of a background wind. Windward, the updraft at the lake breeze front is enhanced and forms a strong vertical plume (Fig. 5j,m), and the inland penetration length is decreased (Fig. 5k,l,n,o). This prevents the cold moist air formed over the lake to spread over land in the upwind direction, which affects the air humidity (Fig. 6d,j,p) and evaporation efficiency (Fig. 6e,k,q) on the upwind side of the lake. Leeward, the local surface wind of the lake breeze is enhanced by the background wind going in the same direction (Fig. 5b,c,e,f). A tail of wind acceleration forms behind the lake, extending >200 km away from the lake center. The horizontal wind at altitude (from around 30 m to several hundred meters) is accelerated to a value higher than the sum of the lake breeze without background wind and the background wind (e.g., to 0.2 m/s at 200 m in altitude, while the background wind there would be ∼0.05 m/s, and the lake breeze alone <0.1 m/s; see Fig. 6b,n). This accelerated wind pocket at altitude forms essentially at night (see Fig. 5b,e,h,k,n and Fig. 6b) and is not observed in 2D simulations (see Fig. 5l,o). An explanation could be that in 3D the lake breeze is weaker than in 2D, especially at night, and that the upper return branch of the lake breeze leeward is not strong enough to form against the background wind. Strong winds coming from above can then join the return branch of the windward lake breeze cell and blow strongly leeward (see Fig. 5k). Convergence of the upper winds coming from the semi-circular front of the lake breeze also enhances the acceleration of the wind locally just behind the lake. As a result of the combination of the eastward lake breeze and the eastward background wind, the lake breeze on the downwind side does not form a clear front, and there is no updraft either. On this side, the evaporated methane vapor is advected over the land more efficiently, leading to drier conditions at the shore, but with more evaporation and colder temperatures over the lake.

In conclusion, in Titan's conditions at 74°N at the autumn equinox, while the mean background wind used in the simulations is not strong enough to suppress lake breezes, it strongly modifies their structure. A strong acceleration of the wind behind the lake at a few 10s to 100 s of meters in altitude is predicted. 3D simulations are necessary to study this phenomenon since it is not present in 2D simulations.

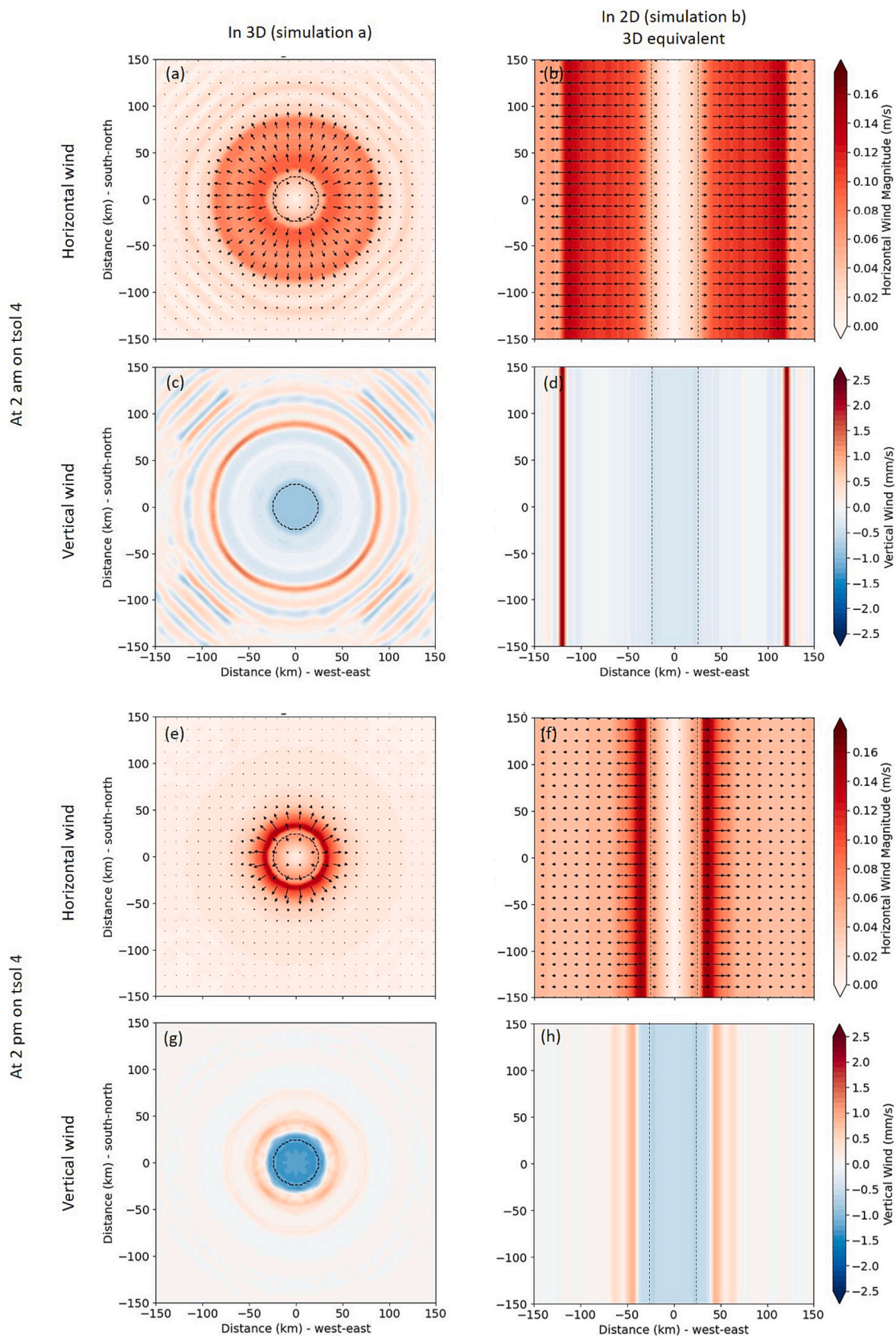

**Fig. 3.** Maps of horizontal (at 2.6 m) and vertical (at 170 m) winds at 2 am and 2 pm local time. Comparison of atmospheric circulations obtained in 3D (simulation a) and 2D (simulation b) for a 50 km diameter lake. In the 2D case, the equivalent 3D map is plotted, with a lake infinitely long along the y-axis. The black dashed lines indicate the position of the lake. The arrows indicate the direction and relative intensity of the horizontal wind at 2.6 m. Note: the quasi-periodic features seen on panels (a) and (c) are gravity waves created by the lake breeze front that resonate in the simulation box in the very calm night conditions. On Titan, such waves would generally not resonate, unless very specific lake configurations.

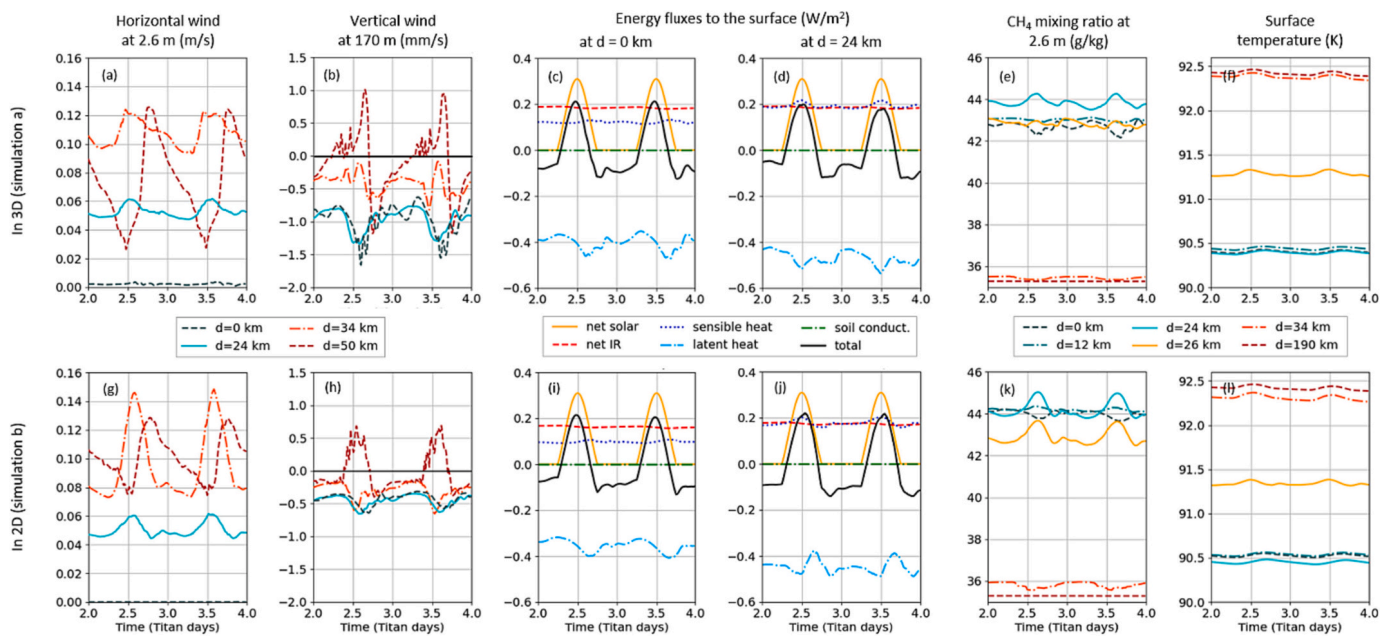

**Fig. 4.** Evolution with time of surface horizontal wind, vertical wind at 170 m, energy fluxes over the lake, methane mixing ratio and surface temperature. Comparison of results obtained in 3D (simulation a) and 2D (simulation b) for a 50 km lake. The parameter d is the distance to the center of the lake. Energy fluxes are defined from the atmosphere (or from the subsurface) to the surface so that positive fluxes participate to heat the surface, and negative fluxes to cool the surface.

## 4. Influence of shoreline: examples with some Titan's lakes

### 4.1. Influence of shoreline concavity

As mentioned in Section 3.1, the curvature of the shore affects the divergence of the lake breeze at the shore. Here we explore curvature effects by studying how a realistic shoreline (Hayes, 2016) affects a lake breeze circulation. We simulated, with and without background winds, the lake breeze above two lakes of irregular shapes at the same latitude (74°N) on Titan: Feia Lacus, which has two inlets (narrow strips of lake that go into the land) creating a peninsula (simulations c and C), and Abaya Lacus, which consists of two lakes touching in one point (simulations d and D). Maps comparing winds, evaporation, temperatures and humidity for the round lake, Feia and Abaya are displayed in Fig. 7 (without background wind) and in Fig. 8 (with background wind). Far from the lake, beyond approximately more than one lake radius from the shore, the lake breeze circulations of Feia and Abaya are quite similar to the one of the round lake. However, as expected, close to the shore and above the lake, the circulation is constrained by the lake shape.

Just above the lake, surface winds increase with the distance from the center of the lake, and are stronger at the extremities of inlets. Stronger surface winds at inlet extremities enhance the evaporation efficiency, as seen with the latent heat flux (e.g. see Feia Lacus on Fig. 7, lines 3 and 4 and column 4), which varies horizontally between 0.4 and 1.1 W/m$^2$. Due to evaporative cooling, the lake is also colder at the extremities of inlets (−0.4 K; Fig. 7 l.5 col. 4). These horizontal variations of the lake evaporation and temperature could lead to horizontal compositional and temperature gradients in the lake that might trigger liquid circulation and mixing (which are not simulated in this work). The combination of asymmetric surface winds and heterogeneous evaporation creates a heterogeneous distribution of cold moist air at the surface of the lake (Fig. 7 l.6,7 col. 4). Consequently, at the surface the cold moist air is especially located over inlets and lake extremities. In these simulations done over small lakes at a latitude of 74°N and at the autumn equinox, the maximum relative humidity reaches 86% at 2.6 m (the lowest simulated level) over inlets and lake extremities (Fig. 7 l.7). Saturation conditions are not reached, but slightly different seasonal, latitudinal or topographic conditions could possibly allow the formation of low clouds or fog. The strong vertical gradient in humidity in the first levels (Fig. 7, l.7,9) is due to the extremely low turbulence above the lake (discussed in more detail in Section 6.2).

The horizontal winds over the land close to lakes with complex shapes are also more variable in intensity and direction compared to the winds around circular lakes (see Fig. 7 l.2,3). Just like on Earth (Crosman and Horel, 2010; McPherson, 1970), peninsulas have a convex shoreline that drives lake breeze winds to converge over the land, and, consequently, to form a strong wind channel along the peninsula flowing from the lake to the land. This does not happen above concave shorelines where lake breeze winds diverge at the shore. Also, the strong wind channels above the peninsulas extend from the surface to a few hundred meters altitude (from 200 to 500 m depending on the insolation; see Fig. 7 l.2 col.3–6). These winds are unique to peninsulas. In the case of round lakes and concave shorelines there is no horizontal wind at 200 m altitude above the shore (Fig. 7 l.2 col.1–2), where the central subsidence of the lake breeze circulation dominates. While the subsidence is homogeneous over the lake in the round case, it is not with complex shorelines (see Fig. 7 l.1). The inlet-shaped sections of the lake evaporate more methane (see Fig. 7 l.4 col.1,2 to compare to col.3–4), and the resulting methane-rich air locally overrides the lake breeze central subsidence and forms small plumes up to a few hundreds of meters in height (Fig. 7 l.1 col.3–4). The cold moist air thus transported within the plumes above the inlets then feeds the peninsula wind channels (seen in the air temperature and humidity maps at 200 m on Fig. 7).

Fig. 8 shows how the lake breeze structure is affected by both the convexity of the shoreline and the presence of a background wind. These two effects interfere with each other, creating unique wind conditions over the lake and the shore. In particular, the peninsula winds seen in Fig. 8 l.2–3 (without wind) and the pocket of accelerated wind downwind seen in Fig. 9 l.2–3 col.1–2 (with a round lake) are both deformed in the case of a lake with a complex shape and background wind (Fig. 9 l.2–3 col.3–6).

### 4.2. Interference of lake breeze circulations created by several lakes

In the *lake district* on Titan (MacKenzie et al., 2019a), many lakes are concentrated in the same region. Previous work in regions with several lakes on Earth (Daggupaty, 2001) noted that lake breezes induced by several lakes could interfere, limit their respective extension, and create strong convective activity at the confluence of the lake breeze fronts. To investigate how the proximity to other lakes could influence the circulation above one lake in Titan's *lake district*, we simulated the lake breeze of Oneida Lacus as if it were alone (simulations e and E) and when surrounded by its nearby lakes (simulations f and F). Results of the wind circulation are given in Fig. 9 in the case without background wind and in Fig. 10 in the case with background wind.

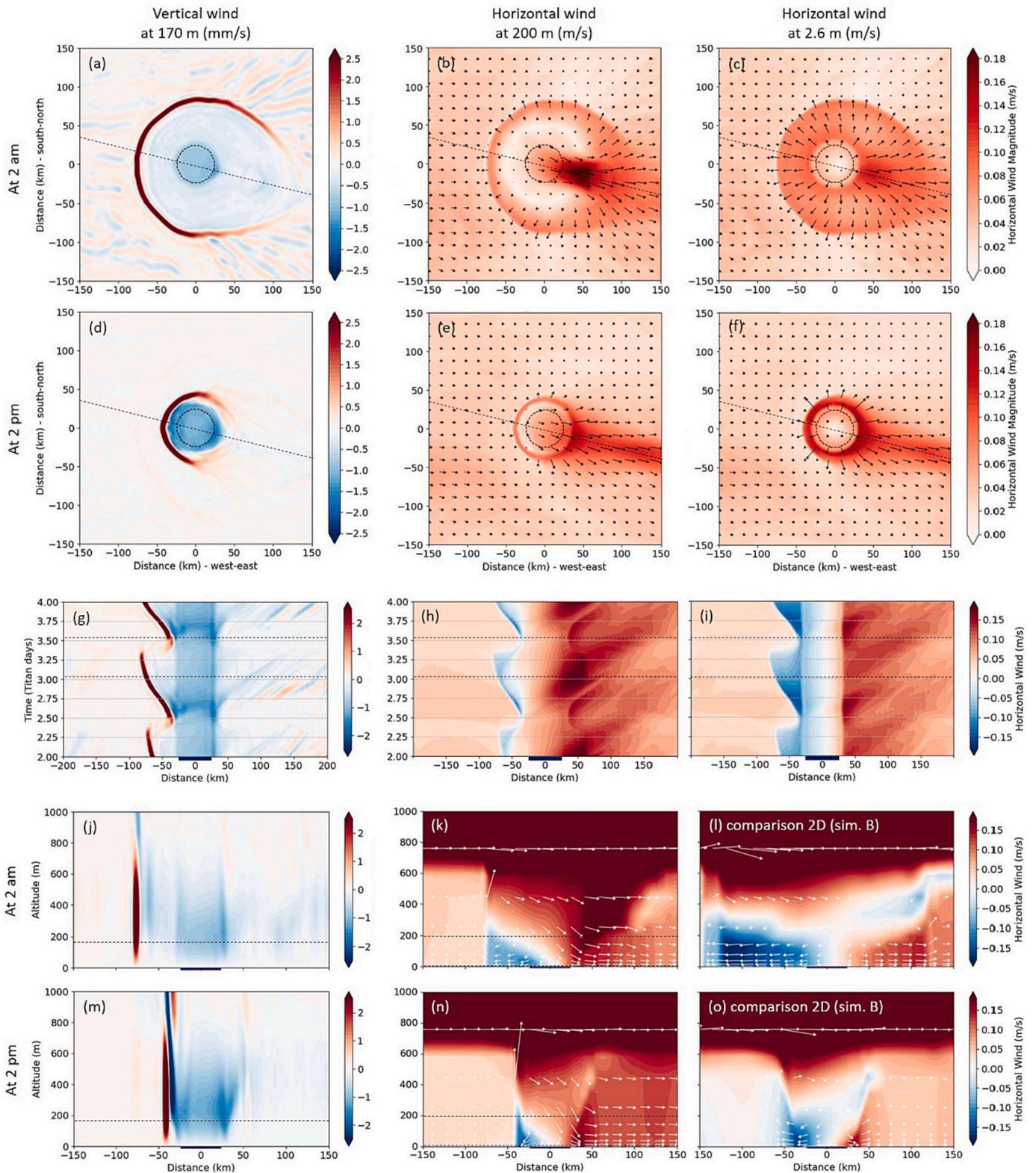

**Fig. 5.** Horizontal and vertical winds on tsol 4. (a–k,m–n) Horizontal and vertical winds for simulation A, with a round lake of 50 km with GCM background winds. The black dashed circles (a–f) and blue line bars (g–o) indicate the position and size of the lake. The arrows indicate the direction and relative intensity of the wind. For (k,l,n,o) the vertical wind is ×100 compared to the horizontal wind for sake of clarity. The dashed black lines on panels (a–f) indicate the position of the slices plotted with different axes on panels (g–o). Dashed black lines on panels (g–i) show the times in which panels (a–f) are plotted. Dashed black lines on panels (j, m) show the 170 m altitude level. Dashed black lines on panels (k) and (n) show the 2.8 and 200 m altitude levels. Panels (l) and (o) give wind results in the 2D case for comparison (simulation B). (For interpretation of the references to colour in this figure legend, the reader is referred to the web version of this article.)

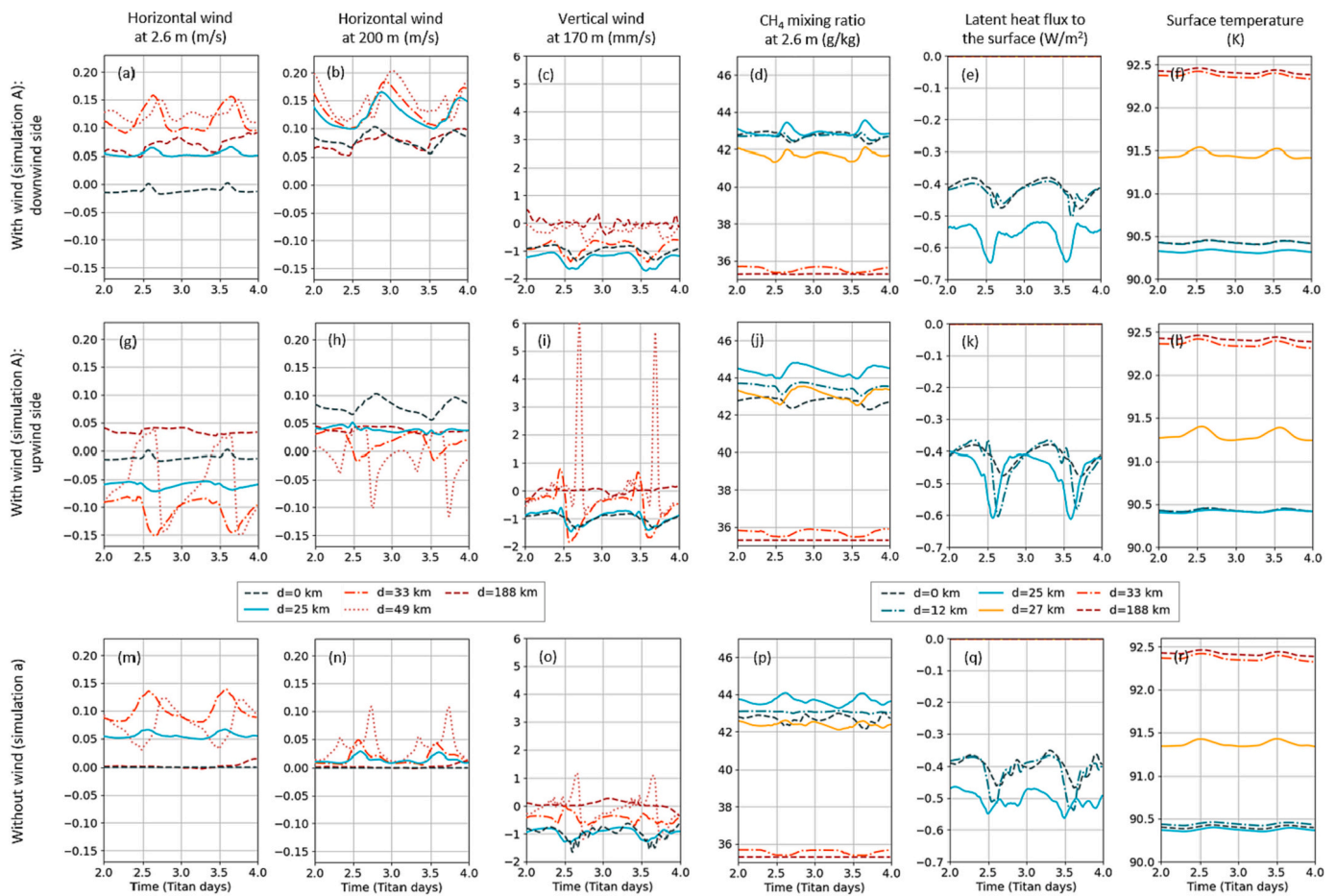

**Fig. 6.** Evolution with time of horizontal wind at 2.6 m and 200 m, vertical wind at 170 m, methane mixing ratio at 2.6 m, latent heat flux to the surface and surface temperature. Comparison of results obtained on the upwind (top row) and downwind sides (middle row) of a simulation with background wind (simulation A) and without background wind (bottom row, simulation a) for a lake of 50 km. d is the distance to the center of the lake, along the slice shown with a dashed line on Fig. 5a.

The main result from these simulations is that the presence of nearby lakes strongly affects the resulting circulation, which is not the linear sum of the lake breeze circulations of all the lakes taken independently. Depending on the local time and the extent of the circulation around the lake, several lake breeze circulation cells can develop (e.g., at 2 pm; see Fig. 9 l.4 col.1), or only a single large breeze circulation cell but including all the lakes (e.g., at 2 am; see Fig. 9 l.2 col.1). On Earth, we usually observe the case of several lake breeze circulations in regions with several lakes (Daggupaty, 2001). On Titan, our simulations show that if two lakes that have independent lake breeze circulations are close enough to each other (e.g., for the conditions at 2 pm, separated by less than the lake size), the individual extents of the lake breezes are shortened, the horizontal wind becomes nearly zero at the region of convergence, and strong upward vertical winds form (Fig. 9 col.1 l.4 to be compared to l.3). In contrast, at night there is a single lake breeze cell that includes all the lakes in a large subsidence zone (Fig. 9 col.1 l.2).

In both cases, though more strongly in the single cell case, zones surrounded on one side by several lakes act like large peninsulas where lake breeze winds converge and create a strong current flowing from the central lake to the land (Fig. 9 col.2–3 l.2,4). Like the peninsula currents in the single lake case, these currents extend higher than the usual lower branch of the lake breeze, up to 800 m (not shown). This affects the upper return branch of the lake breeze, which usually stands around 400–600 m (Fig. 2 a,e). As a result, the upper return branch of the lake breeze is strengthened and starts at lower altitude (above 150 m) between the peninsulas (that is to say, above the lateral lakes, as seen at 200 m in Fig. 9 col.2 l.2,4). Consequently, winds are stronger and of completely different directions to the case with a single lake.

Fig. 10 shows the resulting simulations with an input background wind, in the case with Oneida Lacus alone (simulation E) and Oneida Lacus surrounded by other lakes (simulation F). Therefore, it shows the combined effects of the background wind, the use of real lake shapes, and the presence of nearby lakes. The resulting circulation is complex, with several circulation cells (e.g. Fig. 10 l.4) or a single one (e.g. Fig. 10 l.2) depending on the local time. Horizontal wind currents can be very different in intensity and direction from the ones observed in the previous simulations (i.e., with only one lake, on Fig. 10 l.1,3, or without background wind, on Fig. 9 l.2,4), especially at a few hundred meters in altitude. To conclude, the wind intensity and direction in a location in the lake district and the resulting air-lake interactions are fundamentally connected to the distribution, shape, and size of all the lakes and to the background wind.

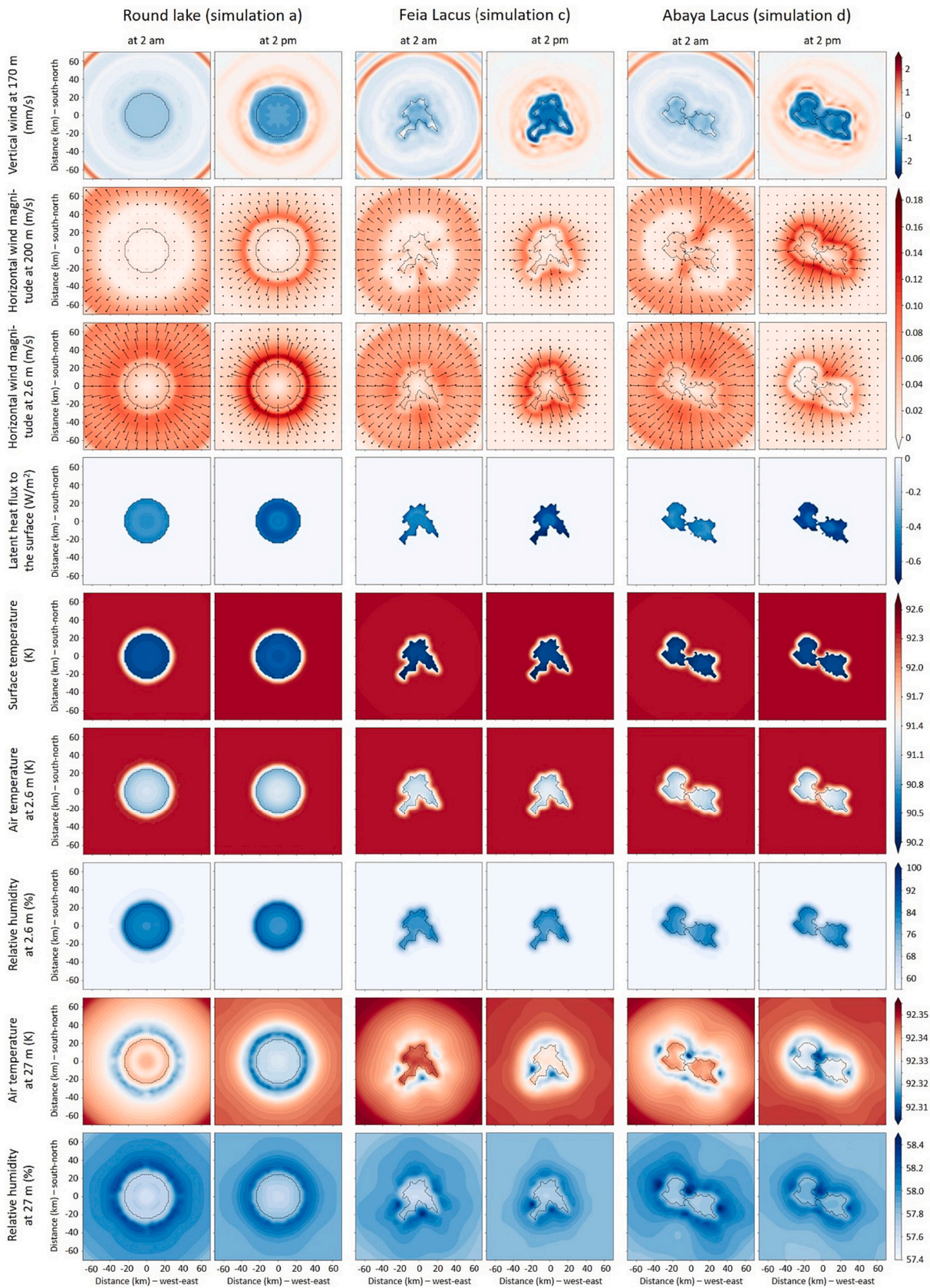

**Fig. 7.** Maps, at 2 am and 2 pm, of vertical wind at 170 m, horizontal wind at 200 m and 2.6 m, latent heat flux to the surface, surface temperature, and air temperature and relative humidity at 2.6 m and 27 m, on tsol 4. Comparison of circulations obtained with a round lake (simulation a), Feia Lacus (simulation c) and Abaya Lacus (simulation d). The black dashed lines indicate the position of the lake. The arrows indicate the direction and relative intensity of the horizontal wind.

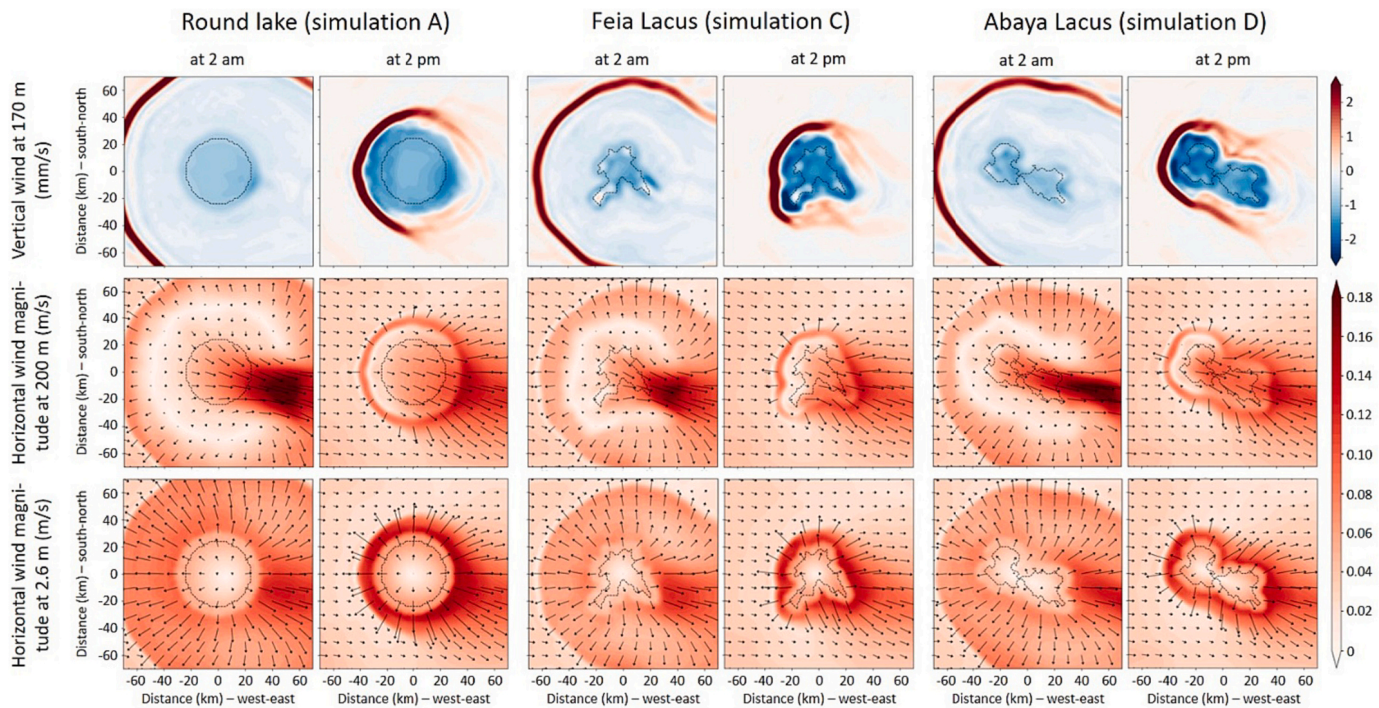

**Fig. 8.** Maps, at 2 am and 2 pm, of vertical wind at 170 m and horizontal wind at 200 m and 2.6 m, on tsol 4. Comparison of circulations obtained with a round lake (simulation A), Feia Lacus (simulation C) and Abaya Lacus (simulation D), with background wind. The black dashed lines indicate the position of the lake. The arrows indicate the direction and relative intensity of the horizontal wind.

## 5. Influence of lake size: examples with some of Titan's lakes

Previous simulations have been run for lakes smaller or close in diameter to 50 km, typical to many Titan's lakes. However, bigger lakes are also found, up to Jingpo Lacus, which is the largest one with a length of 240 km. Previous studies on Earth lakes have investigated the effect of the lake dimensions on the lake breeze (Crosman and Horel, 2010). Those studies observe non-linear variations that combine two effects (Boybeyi and Raman, 1992): i) larger lakes have a larger reservoir of liquid and can therefore evaporate larger amounts of vapor, and ii) larger lakes globally have a smaller curvature, and their lake breeze winds are therefore less divergent. Consequently, larger lakes can form well-developed circulations, with stronger horizontal winds, further inland penetration, stronger and higher updrafts at the fronts, and a weaker central subsidence. Earth studies observe stronger variations with the lake size for small sizes, and much smaller variations with size for lakes larger than 100 km. To quantify the effect of size on the breeze circulation and air-lake interactions on Titan, four simulations of lakes of various lengths with the same horizontal resolution of 4 km, a latitude of 74°N, a domain of 800 × 800 km² and background wind were conducted (Fig. 11 and Fig. 12).

Our simulations indicate that the intensity of the circulation varies with lake size. As seen with Earth lakes, bigger lakes create a lake breeze with stronger horizontal winds in the whole lake breeze structure and in particular in the wind pocket acceleration downstream (e.g., with a wind increased by +25% between simulations G of a round 50-km long lake and J of Jingpo Lacus 240-km long), and a lake breeze front reaching further over land (e.g., a factor of 2 between simulations G and J). By mass conservation, the subsidence over larger lakes is weaker (e. g., a factor of 2 between simulations G and J; Fig. 11 l.1 col.1,4). Surface winds can be locally stronger on larger lakes (e.g., with a maximum surface wind over the lake increased by +40% between simulations G and J; Fig. 11 l.3 col.1,4).

Latent heat flux is mostly spatially homogeneous over the lake centers, but it increases drastically at the shore, from a factor of 2.5 to a factor of 10 depending on the distance to the lake center, the shoreline convexity and the alignment with the background wind (Fig. 12 l.1,4). As the area to perimeter ratio increases for larger lakes, the mean distance of lake points to the shore increases and therefore the mean latent heat flux over the lake decreases (e.g., by −8% between the round 50 and 100 km lakes). Complex shorelines increase the shore length for the same lake area and consequently decrease the mean distance of lake points to the shore. In consequence, the mean latent heat flux slightly increases compared to a smooth circular lake (e.g., by +7% between the round 100 km lake and the similar size lake Bolsena Lacus). The evaporation is closely related to the lake surface temperature due to evaporative cooling. A lower mean latent heat flux in absolute value leads to a higher mean temperature over larger lakes (e.g., +0.15 K between simulations G and J). Humidity depends on the evaporation rate and the local circulation. Zones of higher humidity are formed in larger lakes, on the sections further away from the lake center. For instance, the northern section of Jingpo Lacus, which is roughly 240 km in its longest section, reaches a humidity of 96.3% at 2.6 m above the lake during the night (Fig. 12 l.3 col.4). Condensation could then occur below 2.6 m in these conditions, and possibly also at 2.6 m and above at a different season, as methane evaporation is expected to be stronger during summer (Chatain et al., 2022; Jennings et al., 2019; Lorenz et al., 2012) (see discussion in Section 6.2).

## 6. Discussion

### 6.1. Characteristic dimensions and winds of the lake breeze

A summary of the lake breeze characteristic dimensions and winds for the reference simulation is given in Fig. 13. Our first observation is that while conditions on Titan are extremely different from the ones on Earth, the lake breezes formed around lakes have characteristics within similar orders of magnitude: for the Earth, a wind speed of ~5 m/s, a penetration of ~25 km and a height of ~500 m (Crosman and Horel, 2010), to compare to our Titan simulations with a wind speed of ~0.12 m/s, a penetration of 10–100 km and a height of ~400 m.

A scaling law obtained in Earth studies by Segal et al. (1997) gives an

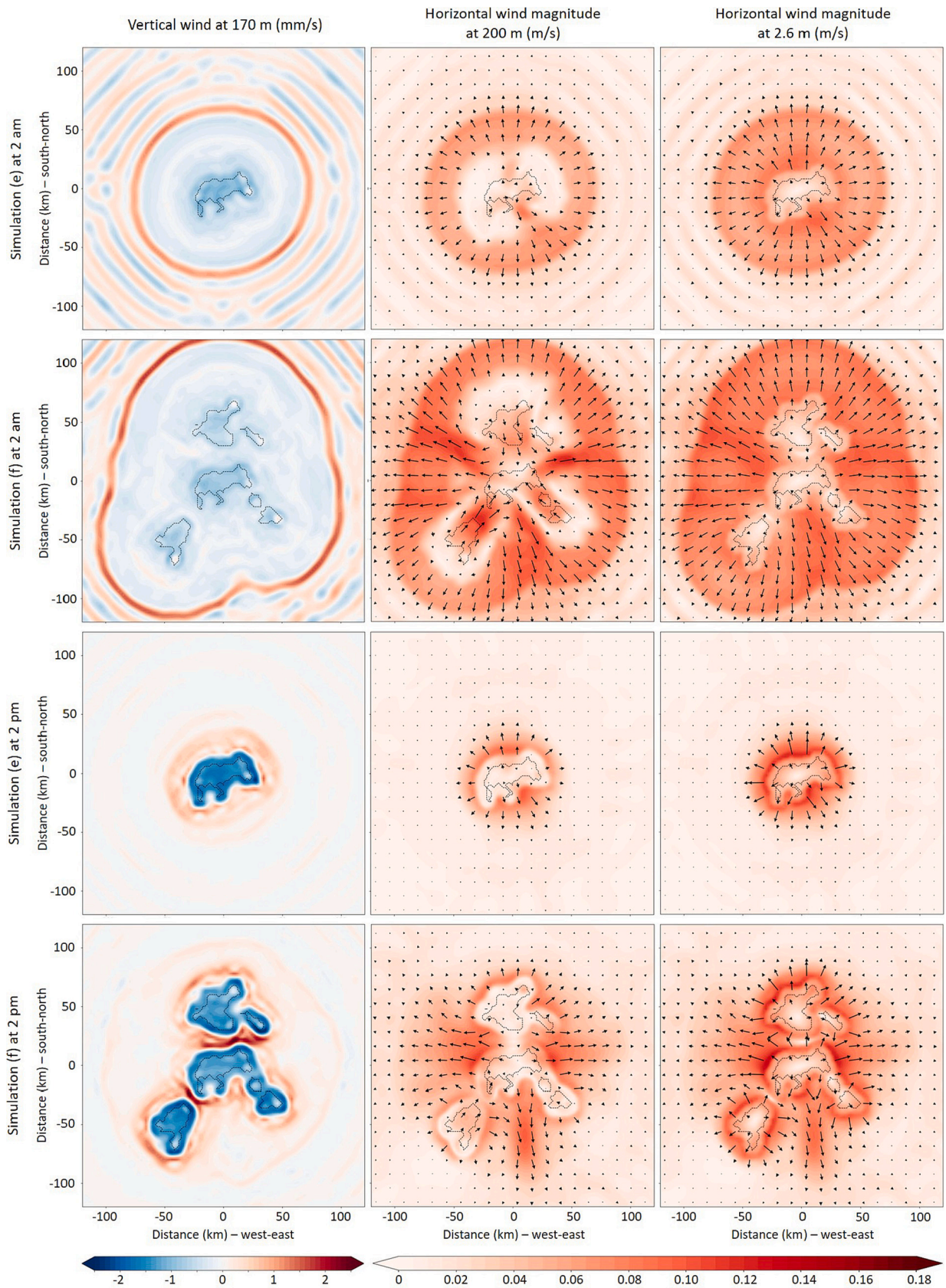

**Fig. 9.** Maps, at 2 am and 2 pm, of vertical wind at 170 m and horizontal wind at 200 m and 2.6 m, on tsol 4. Comparison of circulations obtained with a sole lake (simulation e) and the same lake (Oneida Lacus) surrounded by other lakes (simulation f) without background wind. The black dashed lines indicate the position of the lakes. The arrows indicate the direction and relative intensity of the horizontal wind.

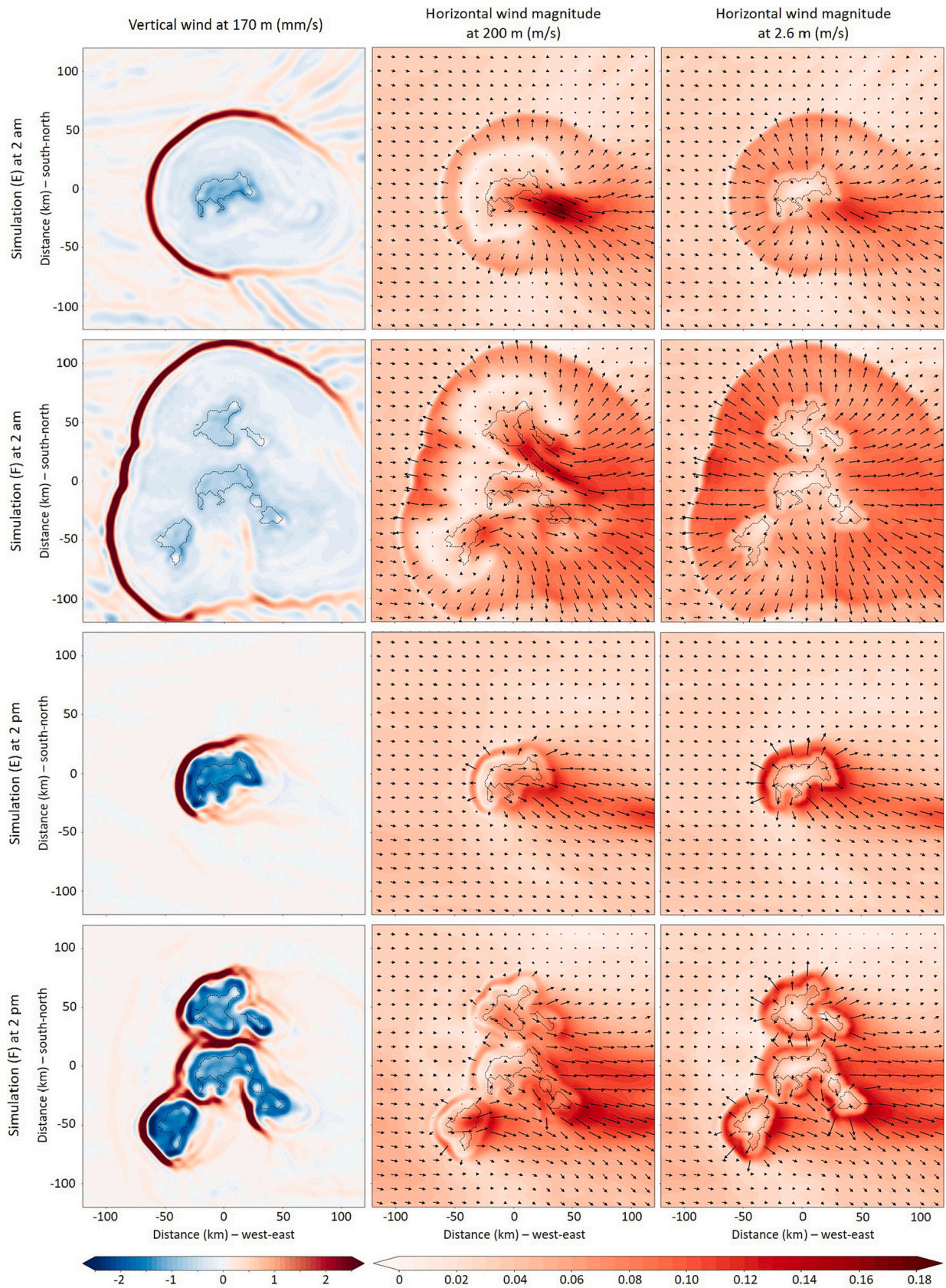

**Fig. 10.** Maps, at 2 am and 2 pm, of vertical wind at 170 m and horizontal wind at 200 m and 2.6 m, on tsol 4. Comparison of circulations obtained with a sole lake (simulation E) and the same lake (Oneida Lacus) surrounded by others (simulation F) with a background wind. The black dashed lines indicate the position of the lakes. The arrows indicate the direction and relative intensity of the horizontal wind.

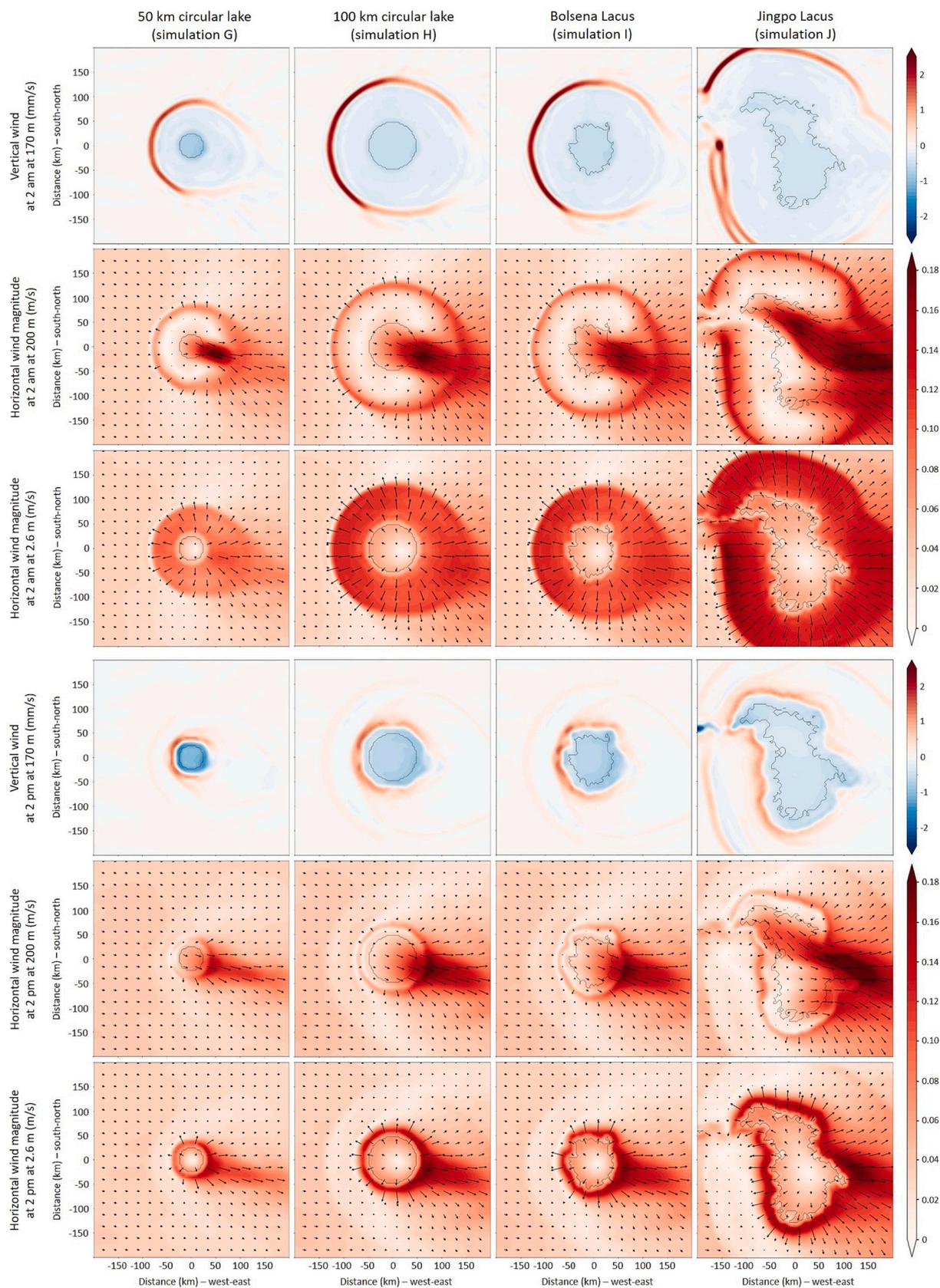

**Fig. 11.** Maps, at 2 am and 2 pm, of vertical wind at 170 m and horizontal wind at 200 m and 2.6 m, on tsol 4. Comparison of circulations obtained with a 50 km circular lake (simulation G), a 100 km circular lake (simulation H), Bolsena Lacus (simulation I) and Jingpo Lacus (simulation J) with a background wind. The black dashed line indicates the position of the lake. The arrows indicate the direction and relative intensity of the horizontal wind.

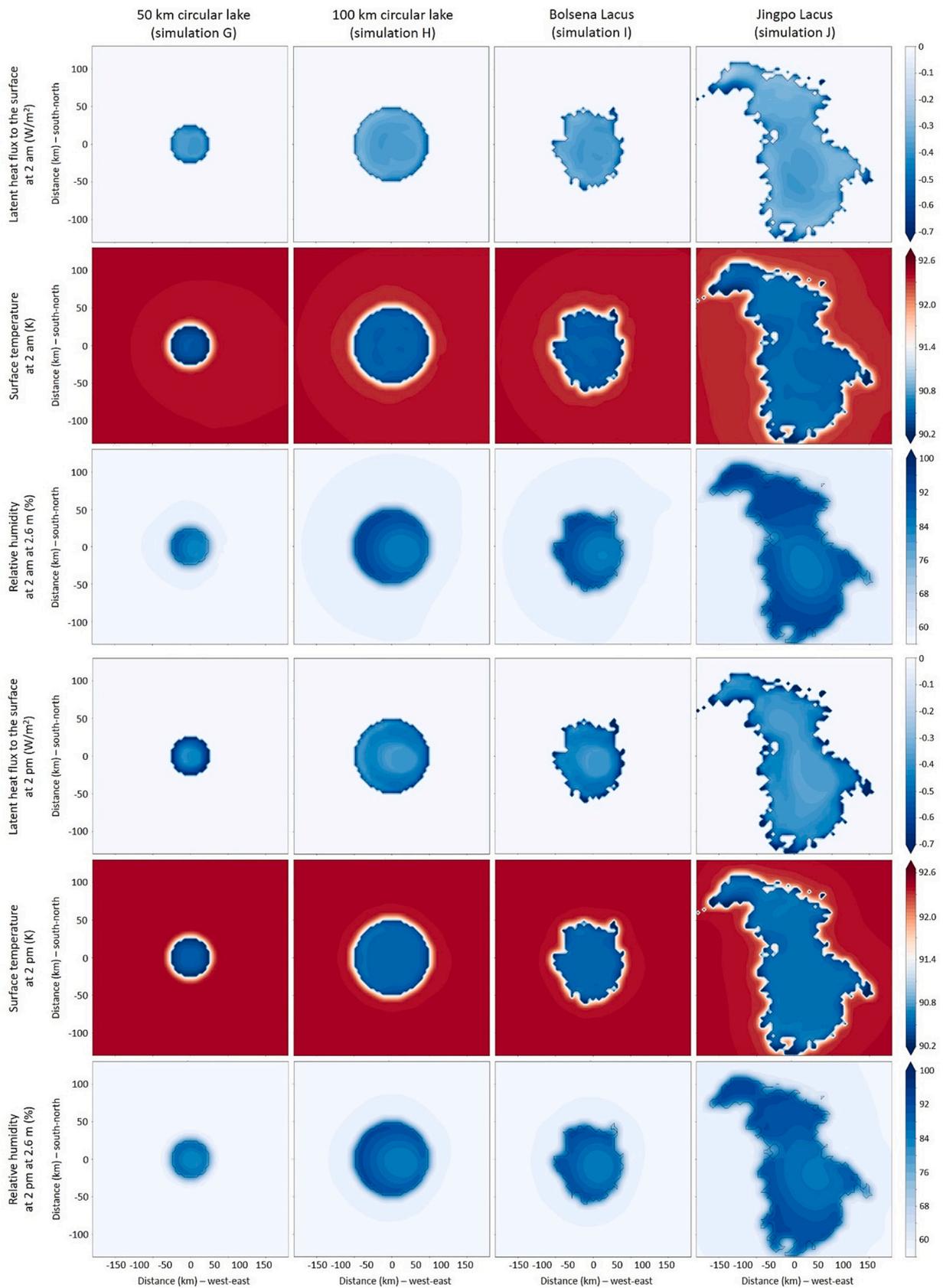

**Fig. 12.** Maps, at 2 am and 2 pm, of latent heat flux to the surface, surface temperature and relative humidity at 2.6 m, on tsol 4. Comparison of circulations obtained with a 50 km circular lake (simulation G), a 100 km circular lake (simulation H), Bolsena Lacus (simulation I) and Jingpo Lacus (simulation J) with a background wind. The black dashed line indicates the position of the lake.

4A. Chatain et al.                                                                                                                                                              Icarus 411 (2024) 115925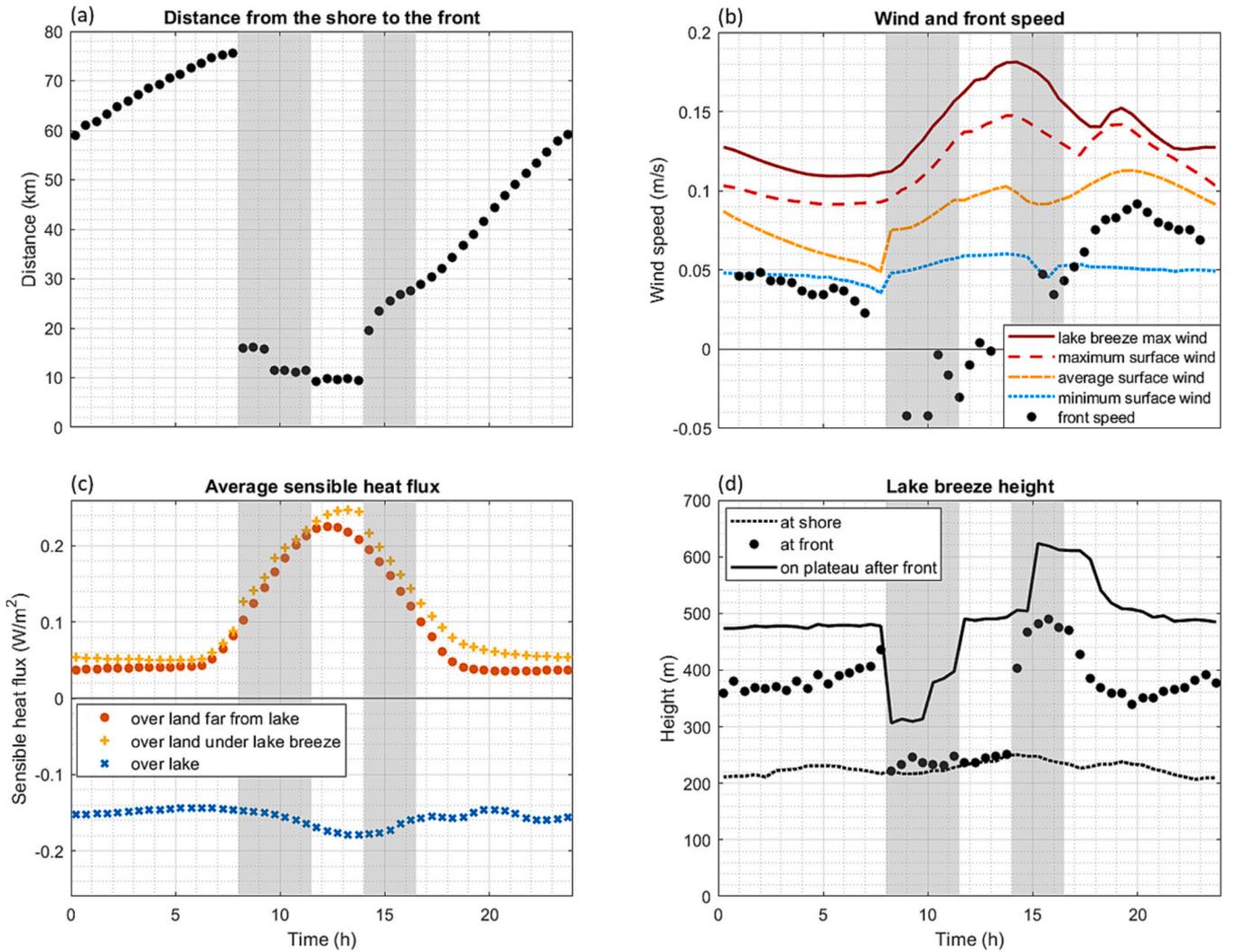

**Fig. 13.** Comparison of the characteristics of the lake breeze in the reference simulation [a] on tsol 4. (a) Distance from the shore to the front, (b) characteristic winds and front speed, (c) sensible heat flux, and (d) lake breeze characteristic heights. The sensible heat flux is spatially averaged over the lake, over the land between the lake and the front, and over the land far from the front. The lake breeze maximum wind is the maximum horizontal wind in the total volume of the lake breeze. The average surface wind is the average of the horizontal wind magnitude in the surface area covered by the lake breeze at the first level (2.6 m). Grey zones indicate times when the front is more diffuse.

idea of the balance of parameters that allows Titan lake breezes to resemble the ones on Earth. The equation is obtained using the assumption that the lake breeze circulation is forced by conversion of potential energy (i.e. temperature) into kinetic energy (i.e. wind) through mass redistribution. Onshore, diabatic heating controls the onshore expansion of the air masses. Offshore, diabatic cooling controls air mass contraction. The horizontal velocity scale $u$ is then expressed as in Eq. (1) (see Segal et al., 1997 for the complete derivation of the equation).

$$u = \left( 1.2 \frac{gHd}{\rho C_p T \left(1 + \frac{d}{l}\right)^2} \right)^{1/3} \quad (1)$$

Comparing Titan to Earth, the gravity acceleration $g$ is smaller by a factor of 7 (from 9.81 to 1.35 m/s$^2$), the land surface sensible heat flux $H$ is smaller by a factor of 1000 (from 300 to 0.2 W/m$^2$), the air density $\rho$ is 5 times larger (from 1.2 to 5.3 kg/m$^3$), the air specific heat capacity $C_p$ does not vary much (from 1005 to 1101 J/kg/K), and the air temperature $T$ is about 3 times lower (from 300 to 92 K). Supposing a similar lake radius $d$ of 25 km and lake breeze length $l$ of 25 km, the equation gives a velocity scale of 4 m/s for the Earth, and 0.16 m/s for Titan. These values are similar to the simulated winds, which suggests that the scaling law [1] taking into account the effects of gravity, sensible heat flux, air density and temperature seems valid for the lake breeze equilibrium conditions on both the Earth and Titan, at least for coarse comparisons.

The lake breeze front expands at a velocity that is of the order of magnitude of the average lake breeze wind (see Fig. 13b). The maximum extension over land can thus be roughly estimated from the horizontal velocity scale, multiplied by the time of extension. Both on Earth and Titan, lake breezes are partly or totally destroyed each day. However, the timing and causes are different. On Earth, lake breezes are best observed on hot summer days when direct insolation quickly increases the land surface temperature. As the lake thermal inertia is higher, the lake surface temperature changes more slowly. A typical horizontal temperature variation of ~5 K then leads to the formation of the lake breeze. As lake breezes form due to solar heating, they last less than half a day on Earth. In our Titan simulations, the temperature difference leading to lake breezes has two sources: the heating of the land through direct insolation during the day, but also the cooling of methane lakes through evaporation. On Titan, these two effects have the same order of magnitude (in particular visible on the sensible heat flux induced by evaporative cooling over the lake, and on the sensible heat flux created by insolation heating over the land, see Fig. 13c), whereas on Earth the insolation process is dominant. The combination of the two processes on Titan leads to a horizontal temperature variation of ~1.2 K between the lake and the land that is roughly constant. The magnitude of this temperature difference is very large in the context of Titan's global surface temperature variations of <4 K (Jennings et al., 2011, 2019). Due to these two processes, lake breezes are present at all hours of the day on Titan. However, during the day the increased subgrid turbulence over the land (see Appendix 1) efficiently mixes the lake breeze front, which is then confined to 10–15 km from the lake shore (see Fig. 13a between 9 am and 3 pm). The lake breeze then starts propagating over land again after 3 pm when the daytime turbulence diminishes. In total, the lake breeze can extend over land during ~18 Titan hours, which is equivalent to 286 Earth hours, some ~50 times longer than the extension time of Earth lake breezes (estimating they extend for ~6 h). Therefore, even if the lake breeze is ~40 times slower on Titan, it can reach a similar or even further distance over land than on Earth.

Fig. 13d shows the evolution of the lake breeze height over one tsol. It is measured as the height at which the horizontal wind becomes zero before reversing direction. The height of the lake breeze above the shore is nearly constant during the day, at around 230 m (see the dotted line). The heights of the breeze at the front (i.e., at the maximum horizontal wind variation; see the black dots) and at a few kilometers after (see the plain line) are also rather constant (at ~400 m and ~ 500 m, respectively; we note that the vertical grid spacing is ~100 m at 400 m), except during daytime when the front is mixed by turbulence. As Titan lake breezes never completely cease, scaling laws made for Earth lake breezes (where the breeze height increases with time starting from a null extension, see Table 2 in Crosman and Horel, 2010) cannot be applied directly. From the parameters used in these laws, the atmospheric stability seems also to play an important role in the breeze height scale, along with air density and surface sensible heat flux. Thus, a future broader investigation of the lake breeze dimensions and winds with different environment conditions (season, latitude, humidity, atmospheric stability, etc.) might help define scaling relations more adapted for Titan.

Fig. 14 compares the lake breeze characteristics found for the various simulations presented in this work. As discussed in the previous sections, it shows that the extension speed of the lake breeze front over land increases with a larger lake (Fig. 14e). The extension speed decreases with a frontal background wind (its speed slows down during the night, Fig. 14d), and in the case of an inlet (Fig. 14f). As a result, during the extension period (i.e., nighttime) the breeze front is further on land with larger lakes, over peninsulas, and without background wind or with a background wind from behind (see Fig. 14a,b).

The plots of maximum winds in the lake breeze and at the surface show a similar pattern in most of the simulations (Fig. 14c,d,e,f): the winds increase during the day, then decrease when the front starts extending over land around 2–3 pm. Around 4–5 pm the winds increase again, and reach a maximum around 7–8 pm before decreasing slowly during the whole night. The second maximum coincides with the front extension maximum speed. We note that the maximum wind in the lake breeze is on average 20% faster than the maximum surface wind. These maximum winds increase with a larger lake (+20% from a lake of 50 km to 100 km), and from an inlet shape to a peninsula shape (+20% in the case of Feia Lacus). The maximum winds reach similar values in the cases without background wind and a frontal background wind. However, the daily variation is slightly different: in the case with frontal background wind, the maximum winds are ~20% slower during midday, but ~20% faster in the evening, with a maximum around 4 pm. Maximum surface winds are found over the land, either at the front or in the wind tail.

## 6.2. Evaporation of the lakes

One of the main goals of the present study is to evaluate the importance of lake evaporation on Titan, to estimate how quickly lake levels could fall because of evaporation, and how much methane could be injected into the atmosphere to contribute to the global methane cycle. The current work provides some estimates toward this. Table 2 summarizes the evaporation rates obtained in all the simulations. The lake level change due to evaporation is computed from the latent heat flux divided by the latent heat of vaporization ($5.1 \times 10^5$ J/kg) and the liquid methane density at Titan's temperature (447 kg/m$^3$).

**Table 2**
Summary of the variations of important parameters observed over the lake in the different simulations.

| mean [min, max] over the lake<br>Simulation | Horizontal wind at 2.6 m (m/s) | Lake level change due to evaporation (cm/Earth year) | Area (km$^2$) | Evaporated mass of CH$_4$ (kg/Earth year) | Relative humidity (%) | Surface temperature (K) |
|---|---|---|---|---|---|---|
| a: circular, 50 km (no background wind) | **0.037**<br>[0.000, 0.067] | 6.1 | 1956 | **5.3 10$^{10}$** | **84.7**<br>[79.7, 89.1] | **90.41**<br>[90.35, 90.46] |
| A: circular, 50 km (with background wind) | **0.039**<br>[0.000, 0.071] | 6.2 | 1956 | **5.4 10$^{10}$** | **85.1**<br>[80.0, 90.1] | **90.41**<br>[90.30, 90.47] |
| B: 50 km large, in 2D | **0.026**<br>[0.000, 0.069] | 5.2 | – | – | **86.7**<br>[83.8, 93.8] | **90.53**<br>[90.44, 90.58] |
| C: Feia Lacus | **0.037**<br>[0.000, 0.076] | 7.3 | 904 | **3.0 10$^{10}$** | **80.6**<br>[71.1, 85.6] | **90.29**<br>[90.00, 90.40] |
| D: Abaya Lacus | **0.033**<br>[0.000, 0.075] | 7.0 | 1164 | **3.7 10$^{10}$** | **80.4**<br>[68.6, 85.3] | **90.32**<br>[90.03, 90.42] |
| E: Oneida Lacus | **0.035**<br>[0.000, 0.074] | 7.4 | 880 | **2.9 10$^{10}$** | **79.6**<br>[65.1, 85.2] | **90.29**<br>[89.88, 90.40] |
| F: Oneida Lacus when surrounded by other lakes | **0.033**<br>[0.001, 0.075] | 7.3 | 880 | **2.9 10$^{10}$** | **79.6**<br>[65.3, 85.7] | **90.29**<br>[89.89, 90.41] |
| G: circular, 50 km | **0.039**<br>[0.002, 0.068] | 6.6 | 2192 | **6.5 10$^{10}$** | **84.0**<br>[77.1, 89.5] | **90.35**<br>[90.14, 90.49] |
| H: circular, 100 km | **0.044**<br>[0.000, 0.076] | 5.6 | 7824 | **19.5 10$^{10}$** | **88.1**<br>[79.6, 93.6] | **90.48**<br>[90.23, 90.57] |
| I: Bolsena Lacus | **0.043**<br>[0.000, 0.080] | 5.9 | 7104 | **18.7 10$^{10}$** | **86.9**<br>[66.8, 93.9] | **90.46**<br>[89.89, 90.58] |
| J: Jingpo Lacus | **0.045**<br>[0.000, 0.096] | 5.7 | 22,720 | **57.8 10$^{10}$** | **87.8**<br>[57.9, 96.3] | **90.50**<br>[89.71, 90.69] |

**Note:** values given in bold are averaged values over the lake on tsol 4. Minimum and maximum values are also given for the wind, the humidity and the surface temperature. Except for the first one (simulation a), cases presented are the ones with background wind, whose values are <5% different from the case without background wind. One season on Titan is equal to 7–8 Earth years.

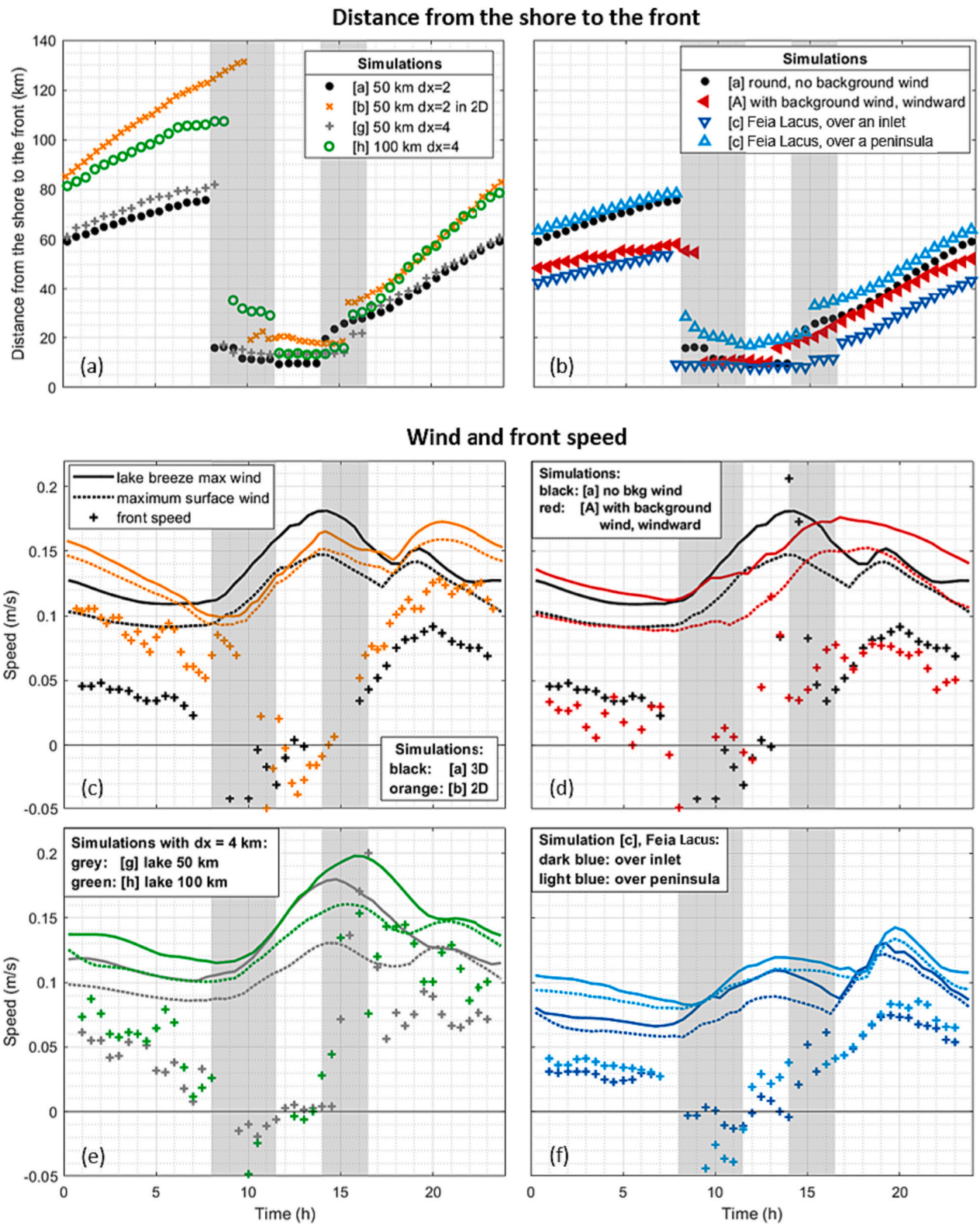

**Fig. 14.** Comparison of the characteristics of the lake breeze in different simulations, on tsol 4. (a,b) Distance from the shore to the front and (c,d,e,f) characteristic winds and front speed. Comparison of simulations a (reference: 50 km round lake without background wind, with a resolution dx = 2 km), A (with background wind), b (in 2D), c (Feia Lacus), g (resolution of dx = 4 km), and h (larger lake). Simulation details are given in Table 1. Grey zones indicate times when the front is more diffuse.

The absolute values should be used with care; while the 3D nature of these simulations is a significant improvement over prior 2D work, the model still lacks other processes, especially internal lake dynamics and surface currents, that may be important. In addition, the absence of methane exchange between the land and the air (Faulk et al., 2020) might drive the model to overestimate the lake breeze intensity, and thus the evaporation rate from the lake. However, if methane exchange between the land and the air were possible, the land would then participate in releasing methane into the atmosphere along with the lake. Also, most lakes on Titan are surrounded by topographic features of several hundred meters (e.g., raised ramparts or rims; Birch et al., 2019; Hayes et al., 2017). These obstacles could substantially affect the winds, and possibly favor the formation of a thicker cold and moist layer above the lakes, which would also decrease the evaporation rate. However, valleys and canyons opening in the lakes would then create strong wind corridors. Further work is needed to take the air-land methane exchange and topography effects into account. Nevertheless, we anticipate that these two effects would on average decrease the lake evaporation rate. Uncertainties on the parametrization of the TKE might also affect the absolute value of these evaporation results by 20–30% (see Appendix 1). Improving these uncertainties would require high resolution Large Eddy Simulations (LES) above a methane lake, or in situ measurements above a lake on Titan (of the turbulence or the methane profile).

The simulations done in this work yield an evaporation rate of ~6 cm per Earth year. This value is significantly lower than previous estimates, and, in particular, much lower than the values of between 20 and 50 cm per Earth year provided in Chatain et al. (2022). This difference is mostly due to the various improvements made to the model, the most dramatic one being the improvement of the turbulence representation. In the new version, the turbulent mixing over the lake is smaller than previously estimated (the Turbulent Kinetic Energy, TKE, is 4 orders of magnitude smaller, see Appendix 1). This allows the formation of a thin and very stable marine layer above the lake, which efficiently reduces the evaporation rate. The decrease of the predicted evaporation rate compared to Chatain et al. (2022) motivates an update on their discussion on the evaporation of lakes (their Section 6.1). The decrease in the expected evaporation rate found in this study may be part of the explanation why no variations are observed in the level of large lakes at the North pole: an underground methane table and some rain could compensate for the low evaporation rate of the lake. However, a slower evaporation rate makes it harder to explain the emptying of small lakes at the North (Hayes et al., 2011; MacKenzie et al., 2019a). Although a more thorough investigation of the evaporation rate at other seasons and latitudes is required, Chatain et al. (2022) suggested that these parameters could only affect the results by less than a factor of 10. Here we support another process that could help empty these lakes: the hydrocarbon liquid from the lakes could percolate through the surrounding solid surface, and possibly also be evaporated at the surface of the land. This process would especially affect lakes surrounded by drier lands (i.e. in the South), or at higher elevations. The seasonal variations of the surface temperature in the north polar region investigated both based on infrared and microwave observations from the Cassini probe (Jennings et al., 2019; Le Gall et al., 2016) indeed suggest the evaporative cooling of the solid terrains surrounding the lakes. We are planning to investigate the efficiency of moist land evaporation in the near future. We expect that land evaporation might decrease the temperature gradient between the land and the lake and thus decrease the lake breeze intensity.

The modification of the turbulent scheme to allow very small values of turbulence over the lakes leads to the accumulation of methane vapor in the first altitude level above the lake, which reaches conditions near saturation. This first model level is at 5 m, and a methane mixing ratio gradient is very likely to be present in these 5 first meters. This suggests that saturation, and so condensation, could occur within the lowest few meters, forming a thin methane fog. The investigation of seasonal effects, the addition of the topography and the inclusion of land moisture in the model might also influence the relative humidity above the lakes and could lead to more favorable fog conditions. These conclusions support the possibility of fog suggested by some VIMS observations both at the South and the North poles (Brown et al., 2009; Dhingra et al., 2020).

We note that the temperature of the lakes does not vary much in our different simulations and compared to our previous study (Chatain et al., 2022). The lake temperature averages ~90.4 K. Lake temperature extremes are found in the case of Jingpo Lacus with background wind: 89.7 K and 90.7 K (see Table 2). This horizontal variation of 1 K within the same lake might induce horizontal currents and vertical overturning within the lake. Lake dynamics could locally heat the surface of the lake and increase the evaporation efficiency. A lake circulation model would thus be needed to better model the lake surface temperature in the case of large lakes. At Titan's North Pole, there are also three large seas, the largest one being Kraken Mare, which is 1170 km-long. At the large scales of seas, the inclusion of differential insolation, the Coriolis force, and the effects of sea currents on the sea temperature are more important than over the lakes. Including such effects in the model will be important to simulate large seas in the future. Existing models of wind-driven and tidal currents show that surface sea currents could be of a few centimeters per second (Tokano, 2010; Tokano and Lorenz, 2015; Vincent et al., 2018). In high current cases, the liquid at the sea surface could be displaced of at most 100 km in 1 to 2 tsols. As such small currents cannot travel through the total length of the seas in a fraction of a tsol, they could not completely homogenize the surface temperature, and sea borders would certainly remain colder than the center. However, currents could locally affect the surface temperature and the evaporation rate, especially near the shore where temperature gradients are the highest.

Surface wind values over the lakes are maximum at the extremities of large lakes, and increase at lower latitudes and during summer (Chatain et al., 2022). The fastest surface winds obtained in the lake breeze in the simulations presented here (at 74° latitude at equinox) are of order 0.1 m/s (see Table 2), which is not enough to form wind-driven waves according to Hayes et al. (2013), who estimate a wind threshold of 0.4 m/s. We plan to add topography to the model in the near future, to estimate how lake surface winds could be affected by cliffs or canyons. Such wind variations would also locally affect the evaporation efficiency.

## 7. Conclusion

Mesoscale simulations above methane lakes on Titan show that a lake-breeze circulation forms around these lakes and remains in place all day long, with some diurnal variations. The characteristic size and speed of the circulation are remarkably close to Earth lake breezes, due to a compensation in the variation of surface planetary parameters such as gravity, sensible heat flux, air density and temperature. Winds are globally ~40 times weaker than on Earth, but the lake-breeze extension time is ~50 times longer (in particular due to longer days). As a result, the breeze front reaches a similar distance from the lake as on Earth.

In this work, we especially focused on the impact of adding a third dimension to our model. We conclude that the use of 2D simulations gives a similar lake breeze circulation as a more realistic 3D simulation, but with an overestimated horizontal extent (by +50%), an overestimated horizontal wind over the land (by a factor of 2), and an underestimated subsidence over the lake (by a factor of 2). These are due to geometrical divergence / convergence effects combined with mass conservation. As a direct consequence from these modifications in the circulation, in 2D the evaporation of the lake is underestimated (by ~15%), the temperature of the lake is overestimated (by ~0.1 K) and the temperature of the land at some distance to the lake is underestimated. 3D simulations with background winds provided by the TAM GCM (Lora et al., 2022) show the formation of a pocket of accelerated wind behind the lake at night, which is not observed in 2D. Therefore, 3D simulations provide a refinement to the 2D solutions that is consequential in the context of Titan's typical temperature and wind structure.

The implementation of the third dimension allowed us to investigate the effect of new parameters such as the concavity of the shoreline and the presence of nearby lakes. Simulations show that horizontal wind channels form from the surface up to 500 m over peninsulas, flowing from the center of the lake to the land, and in the case of several lakes, from the central lake to the land. These wind currents are fueled by small updrafts observed at the extremities of inlets. Depending on the insolation conditions, several nearby lakes can either have individual lake breeze cells whose front convergence forms strong updrafts (at high insolation), or merge into a combined large lake breeze cell (at night). In the large cell configuration, internal wind channels form from external lakes to the central lake. We simulated lake breezes over several known lakes of Titan, and observed that larger lakes create the largest horizontal temperature variations, the strongest winds, and the highest surface humidity locations, getting close to saturation conditions.

The improvement of the definition of turbulent mixing above the lakes led to the formation of a stable cold and moist marine layer in the first 5 m above the lakes, possibly allowing saturation and condensation in the first couple of meters. This layer of high humidity diminishes the evaporation efficiency of lakes, which results in an evaporation rate of ~6 cm/Earth year, notably smaller than previous estimates.

Future in situ missions to Titan's lakes and seas (like proposed in e.g. Oleson et al., 2015; Stofan et al., 2013) would bring invaluable data to complete the tuning of the model, validate its results and improve our understanding of the air-lake interactions on Titan. Most importantly, largest model uncertainties concern the air mixing in the lowest 10 m above the lake. They could be solved by measurements of the small-scale turbulence (like on Mars with a high frequency pressure sensor, Chatain et al., 2021) and/or the altitude-resolved methane humidity above the lakes. To validate the model results, it would also be interesting to measure the cloud opacity in the lowest 10 m (to look for fog), the winds on the lake and the shores, the lake surface temperature, and the lake surface mixing layer. It would be valuable to capture the diurnal (and if possible the seasonal) variations of these variables, with for example samplings every Titan hour during at least one tsol.

### CRediT authorship contribution statement

**Audrey Chatain:** Conceptualization, Data curation, Formal analysis, Funding acquisition, Investigation, Visualization, Writing – original draft, Writing – review & editing. **Scot C.R. Rafkin:** Methodology, Resources, Software, Supervision, Validation, Writing – review & editing. **Alejandro Soto:** Software, Supervision, Validation, Writing – review & editing. **Enora Moisan:** Investigation, Visualization, Writing – review & editing. **Juan M. Lora:** Resources, Writing – review & editing. **Alice Le Gall:** Resources, Writing – review & editing. **Ricardo Hueso:** Project administration, Supervision, Writing – review & editing. **Aymeric Spiga:** Writing – review & editing.

### Declaration of Competing Interest

The authors declare that they have no known competing financial interests or personal relationships that could have appeared to influence the work reported in this paper.

### Data availability

The Fortran source code of the initialization module developed for this work, the simulation input files, the netCDF simulation outputs, and the post-processing codes to plot figures from the netCDF output files are available on Zenodo with the DOI: 10.5281/zenodo.8172270.

### Acknowledgments

A.C. has received funding for this project from the European Union's Horizon 2020 research and innovation program under the Marie Sklodowska-Curie grant agreement No 101022760. The effort of co-author S.R. was partly supported by the Dragonfly Mission Project at JHU/APL, by institutional overhead, and through a substantial donation of personal time. A. Soto was supported only through institutional overhead and personal time. R.H. was supported by Grant PID2019-109467GBI00 funded by MCIN/AEI/10.13039/501100011033/ and by Grupos Gobierno Vasco IT1366-19. We are grateful to the two unknown reviewers for the addition of interesting discussions to the manuscript.

### Appendix 1. A better turbulence representation for the mtWRF model

We modified the minimum allowable turbulent kinetic energy (TKE) in the Mellor-Yamada-Janjic TKE closure scheme to better represent Titan conditions. In previous works, the minimum value for TKE per unit mass was 0.1 $m^2s^{-2}$, as in the Earth version of the model. While the default value may be appropriate for Earth, it is not appropriate for Titan where the TKE is usually several orders of magnitude lower than on Earth. Use of the default value overrides the predicted TKE value and results in constant parameterized eddy under all simulated conditions. Therefore, we decreased the lower limit to $3 \times 10^{-5}$ $m^2s^{-2}$, to better capture the dynamic range of TKE variations computed in the model. This is consistent with observations: Huygens measurements at the surface of Titan indicated that the turbulence production was 4 orders of magnitude smaller at Titan's surface than on Earth (Tokano et al., 2006).

Fig. A1-1 shows the variation of the TKE with altitude and time in the reference simulation 'a'. It clearly shows the increase of turbulence over the land during the day in the planetary boundary layer. On the contrary, the air above the lake is never turbulent. The front of the lake breeze over land corresponds to a local increase of TKE.

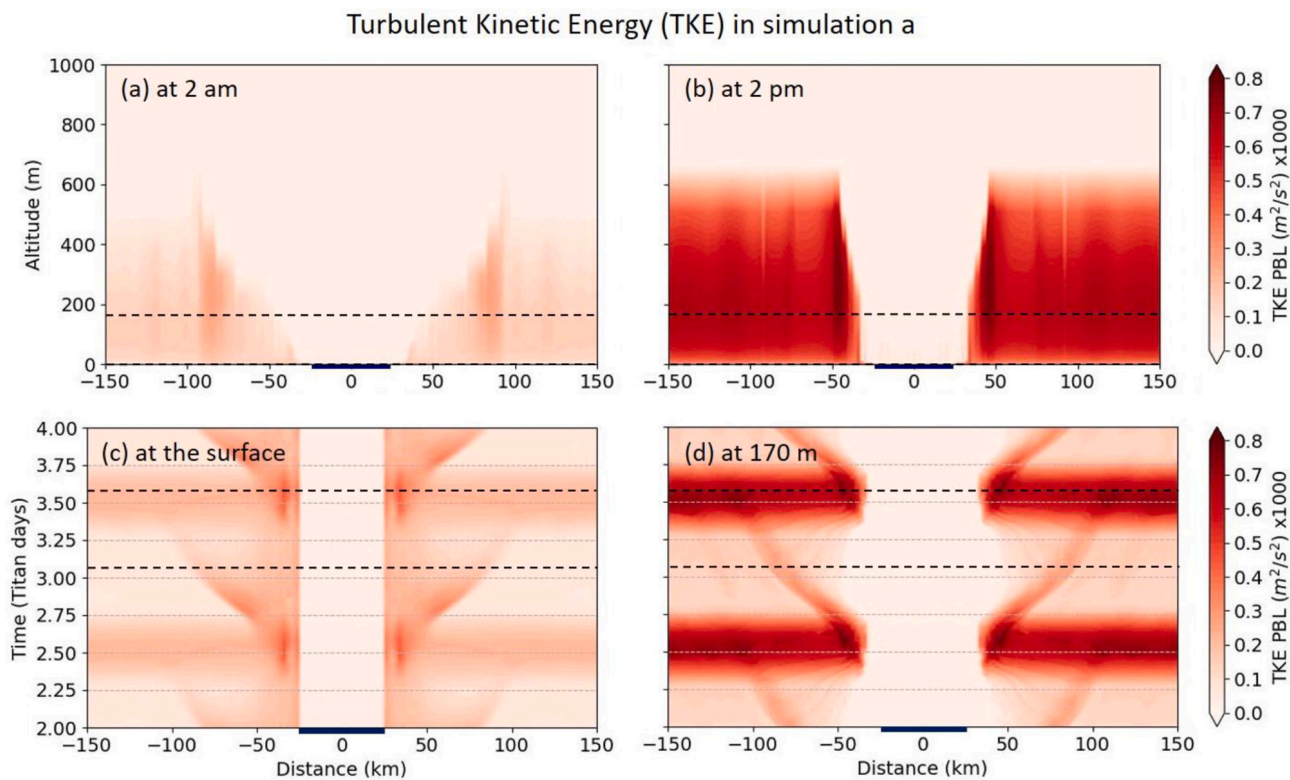

**Fig. A1-1.** Turbulent Kinetic Energy (TKE) as function of altitude and time for a round lake of 50 km without background wind (simulation a). The dark blue bars below the plots indicate the position of the lake. The black dashed lines indicate the position of the slices plotted with different axes. (For interpretation of the references to colour in this figure legend, the reader is referred to the web version of this article.)

The typical TKE above the land is above $2 \times 10^{-4}$ m$^2$s$^{-2}$, which is well above the updated chosen lower limit for TKE. The TKE computed over the lake is much lower, around $3 \times 10^{-6}$ m$^2$s$^{-2}$, due to the more stable conditions. However, $3 \times 10^{-6}$ m$^2$s$^{-2}$ is an extremely low value that would stop all exchanges between the first vertical level (from the surface to 5 m) to the second level. This is not physical and it is a limitation of the vertical grid resolution of mesoscale models to handle extremely low turbulence. Models with higher resolution (called "Large Eddy Simulations") will be necessary to investigate what is happening on the first 5 m over the lake. In the meantime, we have to estimate an equivalent value for the TKE lower limit, to compensate the lack of vertical resolution in the extremely stable zone above the lake, and avoid the total absence of mixing between the two first levels.

Fig. A1-2 compares results obtained for three estimated values for the TKE lower limit: $5\,10^{-5}$ m$^2$.s$^{-2}$ (simulation a2), $3\,10^{-5}$ m$^2$.s$^{-2}$ (simulation a, reference), and $1\,10^{-5}$ m$^2$.s$^{-2}$ (simulation a3). The simulations show that the lake breeze circulation (see rows 1 and 2) and the height of the mixing layer (see row 3) are only slightly affected by the value of the TKE lower limit, which supports the validity of the wind results presented in this paper. The variables the most affected are the one computed at the surface of the lake. A lower TKE leads to a lower mixing, and consequently an accumulation of the methane evaporated by the lake in the first vertical level (i.e. below 5 m, see rows 5 and 6). As the relative humidity in the first level increases, the evaporation efficiency diminishes (see the latent heat flux on row 7). With a lower cooling through evaporation, the final surface temperature increases slightly (of a few tenth of Kelvin, see row 8).

Simulation a2, with a TKE lower limit of $5\,10^{-5}$ m$^2$.s$^{-2}$ has a non-negligible mixing between the two first levels, which allows the methane evaporated from the lake to go up above the first level (see row 5 column 1). Nevertheless, we cannot know if this mixing is overestimated. On the opposite, simulation a3, with a TKE lower limit of $1\,10^{-5}$ m$^2$.s$^{-2}$ has a quasi-inexistent mixing between the two first levels, which creates an unphysically dry layer above the first level (see row 5 column 3). Simulation a, with a TKE lower limit of $3\,10^{-5}$ m$^2$.s$^{-2}$ is the transition between the two previous cases. Without more clues on what should be the equivalent value for the TKE lower limit above the lake with a first vertical level of 5 m, we decided to run our simulations with the central value of $3\,10^{-5}$ m$^2$.s$^{-2}$, to avoid a possible over-mixing (with a value too high), or an unrealistic dry layer (with a value too low).

Plots of the virtual potential temperature (fourth row) show the high stability of the air in the first 400 m above the lake compared to the land (a lower slope means a higher air stability). We note that in the case of no mixing between the two first levels (simulation a3), a small instable layer forms just above the first level (with a vertical slope).

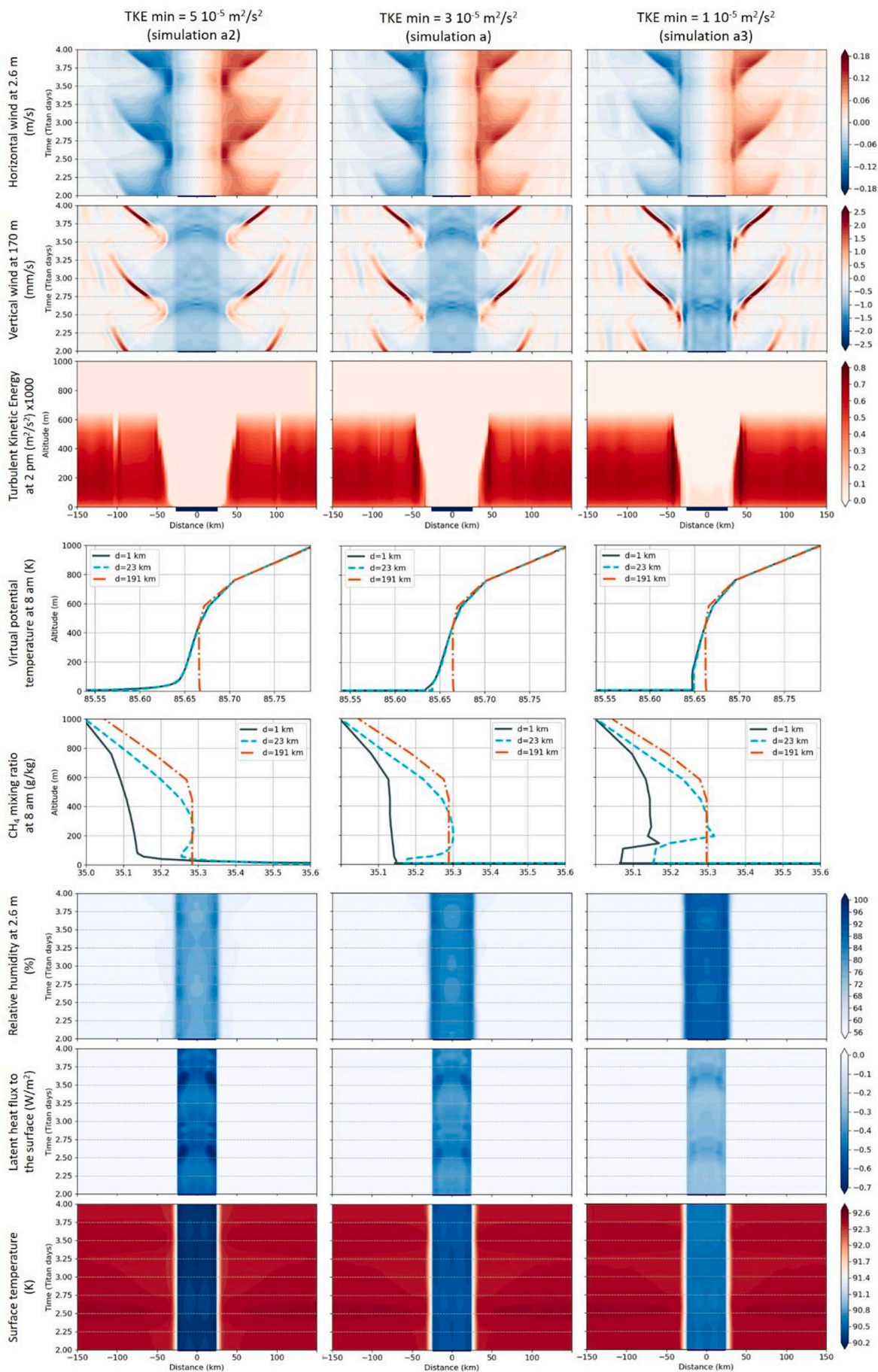

*(caption on next page)*

**Fig. A1-2.** Comparison of simulation results with tree different values for the TKE lower limit: **5 $10^{-5}$ m$^2$.s$^{-2}$ (simulation a2), 3 $10^{-5}$ m$^2$.s$^{-2}$ (simulation a, reference), and 1 $10^{-5}$ m$^2$.s$^{-2}$ (simulation a3).** Horizontal wind at 2.6 m (first row), vertical wind at 170 m (second row), relative humidity at 2.6 m (sixth row), latent heat flux to the surface (seventh row) and surface temperature (eighth row) on a slice crossing the center of the lake, over tsols 3 and 4. Vertical slices of the TKE at 2 pm on tsol 4 (third row). Profiles of virtual potential temperature (fourth row) and methane mixing ratio (fifth row) at 8 am on tsol 4 over the center of the lake (d = 1 km), the lake close to the shore (d = 23 km) and the land far from the lake (d = 191 km). The dark blue bars below the plots indicate the position of the lake. (For interpretation of the references to colour in this figure legend, the reader is referred to the web version of this article.)

**Appendix 2. An updated definition for the underground temperature**

The definition of the subsurface temperature has been improved in this version of the model. As detailed in Chatain et al. (2022), the subsurface temperature is used as a lower boundary condition to compute the soil conduction flux using the soil-slab model described in the appendix of Blackadar (1979). Previously the subsurface temperature was taken as a constant in space (over the land) and time, estimated from the average temperature of the soil over several simulated tsols (Titan days). Theoretically, the subsurface temperature used in this soil-slab model should be the temperature at the bottom of the diurnal wave in the soil, so approximately the average temperature of the surface over several days, which varies with seasons. To make the model more representative of reality, we made two changes. First, the subsurface temperature now depends on the proximity to the lake. It is logically lower closer to the lake whose evaporation cools down the air and persistently cools the nearby ground through sensible heat flux. Second, the subsurface temperature is now computed from the model based on the diurnal evolution of the surface temperature; it is the average value of the surface temperature over the previous 24 h (see Fig. A2). The subsurface temperature and the surface temperature are initialized from a previous simulation in the same configuration, itself initialized with a spatially constant value for the surface and subsurface temperatures estimated from the air temperature at the lowest level of the GCM inputs, and run over 4 tsols.

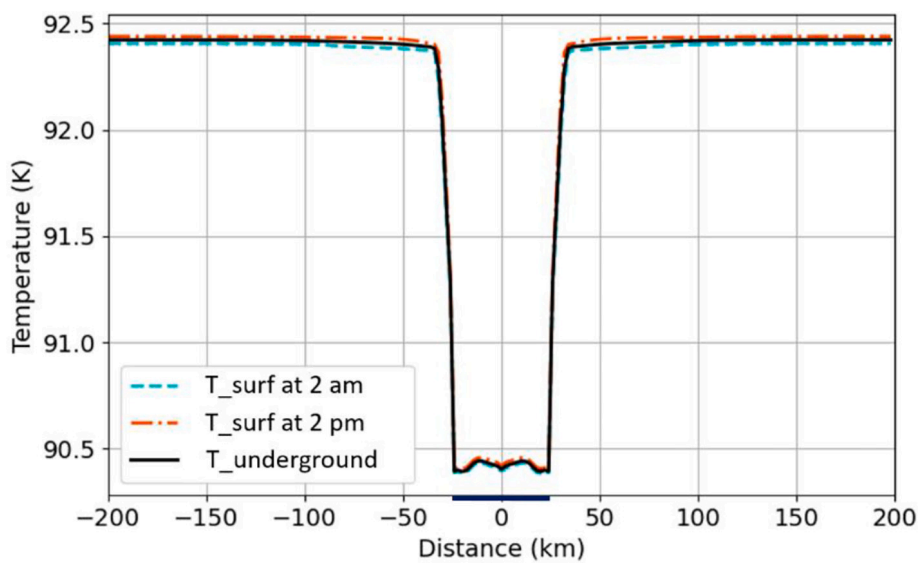

**Fig. A2.** Surface temperature at 2 am and 2 pm, compared to the underground temperature in the case of a round lake of 50 km without background wind (simulation a). The dark blue bar below the plot indicates the position of the lake. (For interpretation of the references to colour in this figure legend, the reader is referred to the web version of this article.)

**Appendix 3. Effect of the horizontal resolution**

In this work, we used two different horizontal resolutions to accommodate either small lakes, large lakes, or groups of lakes. To investigate the robustness of the model results with horizontal resolution, we compared two simulations of a same round lake of 50 km with the two different grid spacings, 2 km (simulation a) and 4 km (simulation g). We observe very similar results on the wind, the latent heat flux, the methane vapor and the surface temperature (see Fig. A3). A few details are worth mentioning. The lake breeze front is ~10 km wide, and consequently the coarser resolution simulation cannot completely reproduce the maximum wind at the front. For this reason, we could also expect slightly stronger winds very locally with a resolution lower than 2 km. Because of a forced 'average' over 4 km in the coarser case, the methane vapor distribution has less spatial variations in simulation g, which in particular smoothens the accumulation of methane vapor at the shore seen in simulation a. The decrease of humidity then leads to an increase of the evaporation at the shore in simulation g, especially during the day. This results in a slightly colder lake at the shore. In conclusion, a finer grid spacing helps resolve the quickly horizontally varying places such as the lake breeze front and the shore. However, the overall shape of the lake breeze and the average intensity of its parameters are reasonably well modelled with a resolution of 4 km. Future simulations with a resolution of 1 km might help defining even more precisely the maximum wind values at the front of the lake breeze (in the order of magnitude of 15%).

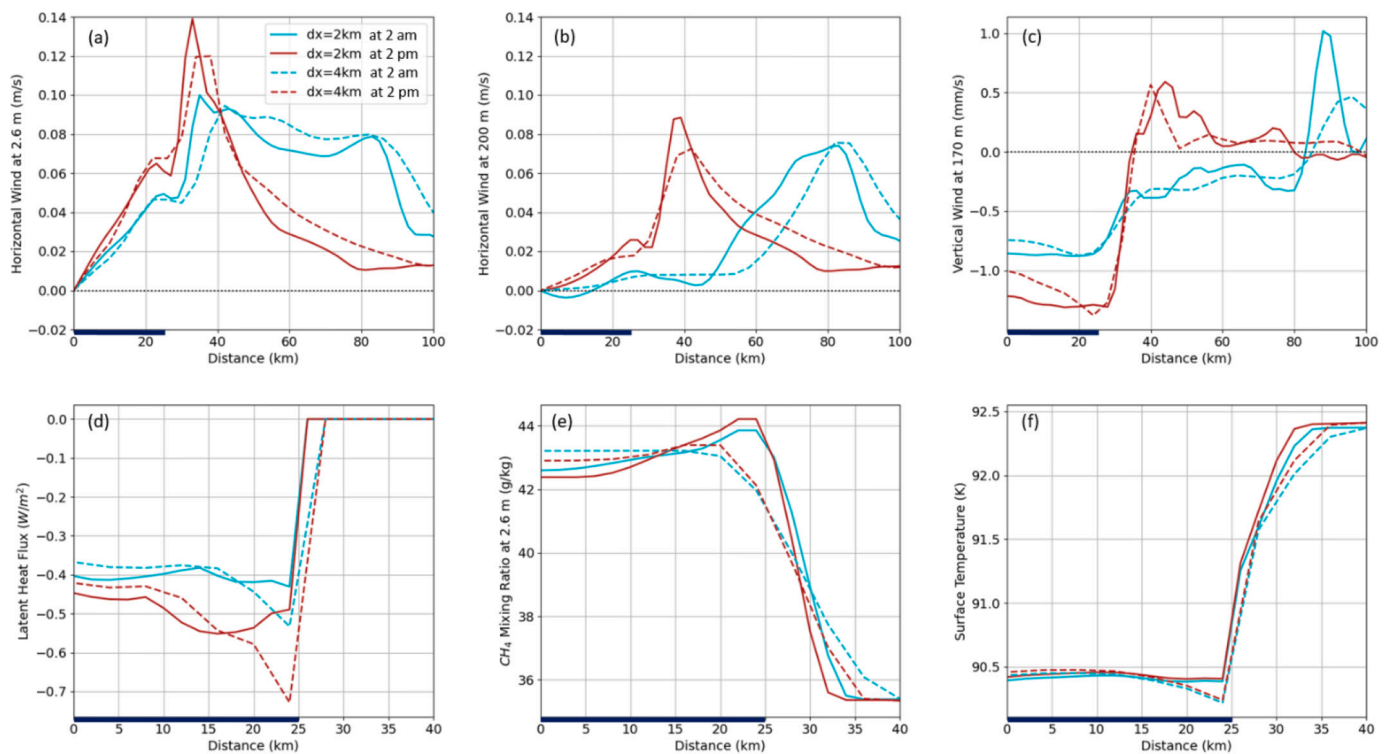

**Fig. A3.** (a) Horizontal wind at 2.6 m, (b) at 200 m, (c) vertical wind at 170 m, (d) latent heat flux to the surface, (e) methane mixing ratio at 2.6 m, and (f) surface temperature, at 2 am (blue) and 2 pm (red). Solid lines are for the reference simulation done with a horizontal resolution dx of 2 km (simulation a). Dashed lines are for dx = 4 km (simulation g). Both simulations have a round lake of 50 km without background wind. The dark blue bars below the plots indicate the position of the lake. (For interpretation of the references to colour in this figure legend, the reader is referred to the web version of this article.)

### References

Arya, P.S., 2001. Introduction to Micrometeorology, Internatio. ed. Academic Press.
Birch, S.P.D., Hayes, A.G., Poggiali, V., Hofgartner, J.D., Lunine, J.I., Malaska, M.J., Wall, S., Lopes, R.M.C., White, O., 2019. Raised rims around Titan's sharp-edged depressions. Geophys. Res. Lett. 46, 5846–5854. https://doi.org/10.1029/2018GL078099.
Blackadar, A.K., 1979. High-resolution models of the planetary boundary layer. Adv. Environ. Sci. Eng. 50–85.
Boybeyi, Z., Raman, S., 1992. A three-dimensional numerical sensitivity study of mesoscale circulations induced by circular lakes. Meteorog. Atmos. Phys. 49, 19–41. https://doi.org/10.1007/BF01025399/METRICS.
Brown, M.E., Smith, A.L., Chen, C., Ádámkovics, M., 2009. Discovery of fog at the south pole of titan. Astrophys. J. 706 https://doi.org/10.1088/0004-637X/706/1/L110.
Changnon Jr., S.A., Jones, D.M.A., 1972. Review of the influences of the Great Lakes on weather. Water Ressour. Res. 8, 360–371. https://doi.org/10.1029/WR008i002p00360.
Chatain, A., Spiga, A., Banfield, D., Forget, F., Murdoch, N., 2021. Seasonal variability of the daytime and nighttime atmospheric turbulence experienced by InSight on Mars. Geophys. Res. Lett. 48, 095453 https://doi.org/10.1029/2021gl095453.
Chatain, A., Rafkin, S.C.R., Soto, A., Hueso, R., Spiga, A., 2022. Air – sea interactions on titan: effect of radiative transfer on the lake evaporation and atmospheric circulation. Planet Sci. J. 3, 232. https://doi.org/10.3847/PSJ/ac8d0b.
Cottini, V., Nixon, C.A., Jennings, D.E., De Kok, R., Teanby, N.A., Irwin, P.G.J., Flasar, F. M., 2012. Spatial and temporal variations in Titan's surface temperatures from Cassini CIRS observations. Planet. Space Sci. 60, 62–71. https://doi.org/10.1016/J.PSS.2011.03.015.
Crosman, E.T., Horel, J.D., 2010. Sea and lake breezes: a review of numerical studies. Bound.-Layer Meteorol. 137, 1–29. https://doi.org/10.1007/S10546-010-9517-9.
Daggupaty, S.M., 2001. A case study of the simultaneous development of multiple lake-breeze fronts with a boundary layer forecast model. J. Appl. Meteorol. 40, 289–311.
Dhingra, R.D., Barnes, J.W., Heslar, M.F., Brown, R.H., Buratti, B.J., Sotin, C., Soderblom, J.M., Rodriguez, S., Le Mouélic, S., Nicholson, P.D., Baines, K.H., Clark, R.N., Jaumann, R., 2020. Spatio-temporal variation of bright ephemeral features on Titan's north pole. Planet. Sci. J. 1, 31. https://doi.org/10.3847/PSJ/ab9c2b.
Drobinski, P., Bastin, S., Dabas, A., Delville, P., Reitebuch, O., 2006. Variability of three-dimensional sea breeze structure in southern France: observations and evaluation of empirical scaling laws. Ann. Geophys. 24, 1783–1799. https://doi.org/10.5194/angeo-24-1783-2006.
Faulk, S.P., Lora, J.M., Mitchell, J.L., Milly, P.C.D., 2020. Titan's climate patterns and surface methane distribution due to the coupling of land hydrology and atmosphere. Nat. Astron. 4, 390–398. https://doi.org/10.1038/s41550-019-0963-0.
Fulchignoni, M., Ferri, F., Angrilli, F., Ball, A.J., Bar-Nun, A., Barucci, M.A., Bettanini, C., Bianchini, G., Borucki, W., Colombatti, G., Coradini, M., Coustenis, A., Debei, S., Falkner, P., Fanti, G., Flamini, E., Gaborit, V., Grard, R., Hamelin, M., Harri, A.M., Hathi, B., Jernej, I., Leese, M.R., Lehto, A., Lion Stoppato, P.F., López-Moreno, J.J., Mäkinen, T., McDonnell, J.A.M., McKay, C.P., Molina-Cuberos, G., Neubauer, F.M., Pirronello, V., Rodrigo, R., Saggin, B., Schwingenschuh, K., Seiff, A., Simões, F., Svedhem, H., Tokano, T., Towner, M.C., Trautner, R., Withers, P., Zarnecki, J.C., 2005. In situ measurements of the physical characteristics of Titan's environment. Nature 438, 785–791. https://doi.org/10.1038/nature04314.
Garratt, J.R., 1977. Aerodynamic Roughness and Mean Monthly Surface Stress over Australia. CSIRO, Melbourne.
Griffith, C.A., McKay, C.P., Ferri, F., 2008. Titan's tropical storms in an evolving atmosphere. Astrophys. J. 687, L41–L44. https://doi.org/10.1086/593117.
Hayes, A.G., 2016. The lakes and seas of titan. Annu. Rev. Earth Planet. Sci. 44, 57–83. https://doi.org/10.1146/annurev-earth-060115-012247.
Hayes, A.G., Aharonson, O., Callahan, P., Elachi, C., Gim, Y., Kirk, R., Lewis, K., Lopes, R., Lorenz, R., Lunine, J., Mitchell, K., Mitri, G., Stofan, E., Wall, S., 2008. Hydrocarbon lakes on Titan: distribution and interaction with a porous regolith. Geophys. Res. Lett. 35, 1–6. https://doi.org/10.1029/2008GL033409.
Hayes, A.G., Aharonson, O., Lunine, J.I., Kirk, R.L., Zebker, H.A., Wye, L.C., Lorenz, R.D., Turtle, E.P., Paillou, P., Mitri, G., Wall, S.D., Stofan, E.R., Mitchell, K.L., Elachi, C., 2011. Transient surface liquid in Titan's polar regions from Cassini. Icarus 211, 655–671. https://doi.org/10.1016/J.ICARUS.2010.08.017.
Hayes, A.G., Lorenz, R.D., Donelan, M.A., Manga, M., Lunine, J.I., Schneider, T., Lamb, M.P., Mitchell, J.M., Fischer, W.W., Graves, S.D., Tolman, H.L., Aharonson, O., Encrenaz, P.J., Ventura, B., Casarano, D., Notarnicola, C., 2013. Wind driven capillary-gravity waves on Titan's lakes: hard to detect or non-existent? Icarus 225, 403–412. https://doi.org/10.1016/J.ICARUS.2013.04.004.
Hayes, A.G., Birch, S.P.D., Dietrich, W.E., Howard, A.D., Kirk, R.L., Poggiali, V., Mastrogiuseppe, M., Michaelides, R.J., Corlies, P.M., Moore, J.M., Malaska, M.J., Mitchell, K.L., Lorenz, R.D., Wood, C.A., 2017. Topographic constraints on the evolution and connectivity of Titan's lacustrine basins. Geophys. Res. Lett. 44, 11,745–11,753. https://doi.org/10.1002/2017GL075468.
Jennings, D.E., Cottini, V., Nixon, C.A., Flasar, F.M., Kunde, V.G., Samuelson, R.E., Romani, P.N., Hesman, B.E., Carlson, R.C., Gorius, N.J.P., Coustenis, A., Tokano, T., 2011. Seasonal changes in Titan's surface temperatures. Astrophys. J. Lett. 737 https://doi.org/10.1088/2041-8205/737/1/L15.
Jennings, D.E., Tokano, T., Cottini, V., Nixon, C.A., Achterberg, R.K., Flasar, F.M., Kunde, V.G., Romani, P.N., Samuelson, R.E., Segura, M.E., Gorius, N.J.P.,

Guandique, E., Kaelberer, M.S., Coustenis, A., 2019. Titan surface temperatures during the Cassini mission. Astrophys. J. Lett. 877, L8. https://doi.org/10.3847/2041-8213/AB1F91.

Kumar, P., Chevrier, V.F., 2020. Solubility of nitrogen in methane, ethane, and mixtures of methane and ethane at titan-like conditions: a molecular dynamics study. ACS Earth Space Chem. 4, 241–248. https://doi.org/10.1021/acsearthspacechem.9b00289.

Le Gall, A., Malaska, M.J., Lorenz, R.D., Janssen, M.A., Tokano, T., Hayes, A.G., Mastrogiuseppe, M., Lunine, J.I., Veyssière, G., Encrenaz, P., Karatekin, O., 2016. Composition, seasonal change, and bathymetry of Ligeia Mare, Titan, derived from its microwave thermal emission. J. Geophys. Res. Planets 121, 233–251. https://doi.org/10.1002/2015JE004920.

Lettau, H., 1969. Note on aerodynamic roughness-parameter estimation on the basis of roughness-element description. J. Appl. Meteorol. 1962-1982 (8), 828–832.

Lopes, R.M.C., Malaska, M.J., Solomonidou, A., Le Gall, A., Janssen, M.A., Neish, C.D., Turtle, E.P., Birch, S.P.D., Hayes, A.G., Radebaugh, J., Coustenis, A., Schoenfeld, A., Stiles, B.W., Kirk, R.L., Mitchell, K.L., Stofan, E.R., Lawrence, K.J., 2016. Nature, distribution, and origin of Titan's Undifferentiated Plains. Icarus 270, 162–182. https://doi.org/10.1016/j.icarus.2015.11.034.

Lora, J.M., Lunine, J.I., Russell, J.L., 2015. GCM simulations of Titan's middle and lower atmosphere and comparison to observations. Icarus 250, 516–528. https://doi.org/10.1016/j.icarus.2014.12.030.

Lora, J.M., Tokano, T., Vatant d'Ollone, J., Lebonnois, S., Lorenz, R.D., 2019. A model intercomparison of Titan's climate and low-latitude environment. Icarus 333, 113–126. https://doi.org/10.1016/j.icarus.2019.05.031.

Lora, J.M., Battalio, J.M., Yap, M., Baciocco, C., 2022. Topographic and orbital forcing of Titan's hydroclimate. Icarus 384, 115095. https://doi.org/10.1016/j.icarus.2022.115095.

Lorenz, R.D., Newman, C.E., Tokano, T., Mitchell, J.L., Charnay, B., Lebonnois, S., Achterberg, R.K., 2012. Formulation of a wind specification for Titan late polar summer exploration. Planet. Space Sci. 70, 73–83. https://doi.org/10.1016/j.pss.2012.05.015.

MacKenzie, S.M., Barnes, J.W., Hofgartner, J.D., Birch, S.P.D., Hedman, M.M., Lucas, A., Rodriguez, S., Turtle, E.P., Sotin, C., 2019a. The case for seasonal surface changes at Titan's lake district. Nat. Astron. 3, 506–510. https://doi.org/10.1038/s41550-018-0687-6.

MacKenzie, S.M., Lora, J.M., Lorenz, R.D., 2019b. A thermal inertia map of Titan. J. Geophys. Res. Planets 124, 1728–1742. https://doi.org/10.1029/2019JE005930.

Mastrogiuseppe, M., Poggiali, V., Hayes, A., Lunine, J., Picardi, G., Seu, R., Flamini, E., Mitri, G., Notarnicola, C., Paillou, P., Zebker, H., 2014. The bathymetry of a Titan Sea. Geophys. Res. Lett. 41, 1432–1437. https://doi.org/10.1002/2013GL058618.

Mastrogiuseppe, M., Hayes, A., Poggiali, V., Seu, R., Lunine, J.I., Hofgartner, J.D., 2016. Radar sounding using the Cassini altimeter: waveform modeling and Monte Carlo approach for data inversion of observations of Titan's seas. IEEE Trans. Geosci. Remote Sens. 54, 5646–5656. https://doi.org/10.1109/TGRS.2016.2563426.

Mastrogiuseppe, M., Poggiali, V., Hayes, A.G., Lunine, J.I., Seu, R., Mitri, G., Lorenz, R. D., 2019. Deep and methane-rich lakes on titan. Nat. Astron. 3, 535–542. https://doi.org/10.1038/s41550-019-0714-2.

McPherson, R.D., 1970. A numerical study of the effect of a coastal irregularity on the sea breeze. J. Appl. Meteorol. 9, 767–777. https://doi.org/10.1175/1520-0450(1970)009<0767:ANSOTE>2.0.CO;2.

Mitchell, J.L., 2008. The drying of Titan's dunes: Titan's methane hydrology and its impact on atmospheric circulation. J. Geophys. Res. Planets 113, 8015. https://doi.org/10.1029/2007JE003017.

Mitchell, J.L., 2012. Titan's transport-driven methane cycle. Astrophys. J. Lett. 756 https://doi.org/10.1088/2041-8205/756/2/L26.

Mitchell, J.L., Lora, J.M., 2016. The climate of Titan. Annu. Rev. Earth Planet. Sci. 44, 353–380. https://doi.org/10.1146/annurev-earth-060115-012428.

Mitchell, K.L., Barmatz, M.B., Jamieson, C.S., Lorenz, R.D., Lunine, J.I., 2015. Laboratory measurements of cryogenic liquid alkane microwave absorptivity and implications for the composition of Ligeia Mare, Titan. Geophys. Res. Lett. 42, 1340–1345. https://doi.org/10.1002/2014GL059475.

Mitri, G., Showman, A.P., Lunine, J.I., Lorenz, R.D., 2007. Hydrocarbon lakes on titan. Icarus 186, 385–394. https://doi.org/10.1016/j.icarus.2006.09.004.

Nelli, N.R., Temimi, M., Fonseca, R.M., Weston, M.J., Thota, M.S., Valappil, V.K., Branch, O., Wulfmeyer, V., Wehbe, Y., Al Hosary, T., Shalaby, A., Al Shamsi, N., Al Naqbi, H., 2020. Impact of roughness length on WRF simulated land-atmosphere interactions over a hyper-arid region. Earth Space Sci. 7, 1–18. https://doi.org/10.1029/2020EA001165.

Newman, C.E., Richardson, M.I., Lian, Y., Lee, C., 2016. Simulating Titan's methane cycle with the TitanWRF general circulation model. Icarus 267. https://doi.org/10.1016/j.icarus.2015.11.028.

Niemann, H.B., Atreya, S.K., Demick, J.E., Gautier, D., Haberman, J.A., Harpold, D.N., Kasprzak, W.T., Lunine, J.I., Owen, T.C., Raulin, F., 2010. Composition of Titan's lower atmosphere and simple surface volatiles as measured by the Cassini-Huygens probe gas chromatograph mass spectrometer experiment. J. Geophys. Res. E Planets 115, E12006. https://doi.org/10.1029/2010JE003659.

Oleson, S.R., Lorenz, R.D., Paul, M.V., 2015. Phase I Final Report: Titan Submarine.

Rafkin, S.C.R., Soto, A., 2020. Air-sea interactions on Titan: lake evaporation, atmospheric circulation, and cloud formation. Icarus 351, 113903. https://doi.org/10.1016/j.icarus.2020.113903.

Schneider, T., Graves, S.D.B., Schaller, E.L., Brown, M.E., 2012. Polar methane accumulation and rainstorms on Titan from simulations of the methane cycle. Nature 481, 58–61. https://doi.org/10.1038/nature10666.

Segal, M., Leuthold, M., Arritt, R.W., Anderson, C., Shen, J., 1997. Small lake daytime breezes: some observational and conceptual evaluations. Bull. Am. Meteorol. Soc. 78, 1135–1147.

Solomonidou, A., Le Gall, A., Malaska, M.J., Birch, S.P.D., Lopes, R.M.C., Coustenis, A., Rodriguez, S., Wall, S.D., Michaelides, R.J., Nasr, M.R., Elachi, C., Hayes, A.G., Soderblom, J.M., Schoenfeld, A.M., Matsoukas, C., Drossart, P., Janssen, M.A., Lawrence, K.J., Witasse, O., Yates, J., Radebaugh, J., 2020. Spectral and emissivity analysis of the raised ramparts around Titan's northern lakes. Icarus 344, 113338. https://doi.org/10.1016/J.ICARUS.2019.05.040.

Steckloff, J.K., Soderblom, J.M., Farnsworth, K.K., Chevrier, V.F., Hanley, J., Soto, A., Groven, J.J., Grundy, W.M., Pearce, L.A., Tegler, S.C., Engle, A., 2020. Stratification dynamics of Titan's lakes via methane evaporation. Planet. Sci. J. 1, 26. https://doi.org/10.3847/PSJ/ab974e.

Stofan, E.R., Elachi, C., Lunine, J.I., Lorenz, R.D., Stiles, B., Mitchell, K.L., Ostro, S., Soderblom, L., Wood, C., Zebker, H., Wall, S., Janssen, M., Kirk, R., Lopes, R., Paganelli, F., Radebaugh, J., Wye, L., Anderson, Y., Allison, M., Boehmer, R., Callahan, P., Encrenaz, P., Flamini, E., Francescetti, G., Gim, Y., Hamilton, G., Hensley, S., Johnson, W.T.K., Kelleher, K., Muhleman, D., Paillou, P., Picardi, G., Posa, F., Roth, L., Seu, R., Shaffer, S., Vetrella, S., West, R., 2007. The lakes of Titan. Nature 445, 61–64. https://doi.org/10.1038/nature05438.

Stofan, E., Lorenz, R., Lunine, J., Bierhaus, E.B., Clark, B., Mahaffy, P.R., Ravine, M., 2013. TiME – The Titan Mare Explorer. In: 2013 IEEE Aerospace Conference. IEEE, Big Sky, MT, USA, pp. 1–10. https://doi.org/10.1109/AERO.2013.6497165.

Tobie, G., Lunine, J.I., Sotin, C., 2006. Episodic outgassing as the origin of atmospheric methane on Titan. Nature 440, 61–64. https://doi.org/10.1038/nature04497.

Tokano, T., 2005. Meteorological assessment of the surface temperatures on Titan: constraints on the surface type. Icarus 173, 222–242. https://doi.org/10.1016/j.icarus.2004.08.019.

Tokano, T., 2010. Simulation of tides in hydrocarbon lakes on Saturn's moon Titan. Ocean Dyn. 60, 803–817. https://doi.org/10.1007/s10236-010-0285-3.

Tokano, T., Lorenz, R.D., 2015. Wind-driven circulation in Titan's seas. J. Geophys. Res. Planets 120, 20–33. https://doi.org/10.1002/2014JE004751.

Tokano, T., Ferri, F., Colombatti, G., Mäkinen, T., Fulchignoni, M., 2006. Titan's planetary boundary layer structure at the Huygens landing site. J. Geophys. Res. Planets 111, 1–10. https://doi.org/10.1029/2006JE002704.

Turtle, E.P., Perry, J.E., Barbara, J.M., Del Genio, A.D., Rodriguez, S., Le Mouélic, S., Sotin, C., Lora, J.M., Faulk, S., Corlies, P., Kelland, J., MacKenzie, S.M., West, R.A., McEwen, A.S., Lunine, J.I., Pitesky, J., Ray, T.L., Roy, M., 2018. Titan's meteorology over the Cassini Mission: evidence for extensive subsurface methane reservoirs. Geophys. Res. Lett. 45, 5320–5328. https://doi.org/10.1029/2018GL078170.

Vincent, D., Karatekin, Ö., Lambrechts, J., Lorenz, R.D., Dehant, V., Deleersnijder, É., 2018. A numerical study of tides in Titan's northern seas, Kraken and Ligeia Maria. Icarus 310, 105–126. https://doi.org/10.1016/j.icarus.2017.12.018.